\documentclass[12pt,a4paper]{article}
\pdfoutput=1
\usepackage{jheppub}
\usepackage[utf8]{inputenc}
\usepackage[T1]{fontenc}

\usepackage{tikz, pgf}   
\usepackage{amsmath,physics,float}  
\usepackage{amssymb}  
\usepackage{latexsym}
\usepackage{graphicx,xcolor}

\usepackage{amsfonts}
\usepackage{braket}
\usepackage{mathrsfs}

\usepackage{subfig} 
\usepackage{empheq}

\usepackage[shortlabels]{enumitem}

\usepackage{pgfplots}
\usepackage{natbib}

\newcommand{\beq}{\begin{equation}}
\newcommand{\eeq}{\end{equation}}

\newcommand{\ksi}{\ket{\psi}}
\newcommand{\del}{\partial}
\newcommand{\lc}{\left(}
\newcommand{\rc}{\right)}
\newcommand{\ls}{\left[}
\newcommand{\rs}{\right]}

\begin{document}

\title{\boldmath Overcounting of interior excitations: A resolution to the bags of gold paradox in AdS}
\author{Joydeep Chakravarty}

\emailAdd{joydeep.chakravarty@icts.res.in}

\affiliation{International Centre for Theoretical Sciences (ICTS-TIFR),
Tata Institute of Fundamental Research,
Shivakote, Hesaraghatta,
Bangalore 560089, India.}
\begin{abstract}
{In this work, we investigate how single-sided and eternal black holes in AdS can host an enormous number of semiclassical excitations in their interior, which is seemingly not reflected in the Bekenstein Hawking entropy. In addition to the paradox in the entropy, we argue that the treatment of such excitations using effective field theory also violates black holes' expected spectral properties. We propose that these mysteries are resolved because apparently orthogonal semiclassical bulk excitations have small inner products between them; and consequently, a vast number of semiclassical excitations can be constructed using the Hilbert space which describes black hole's interior. We show that there is no paradox in the dual CFT description and comment upon the initial bulk state, which leads to the paradox. Further, we demonstrate our proposed resolution in the context of small $N$ toy matrix models, where we model the construction of these large number of excitations. We conclude by discussing why this resolution is special to black holes.}
\end{abstract}

\maketitle

\section{Introduction}
\label{intro}

The Bekenstein-Hawking entropy is a thermodynamic coarse-grained measure which states that the entropy of the black hole is proportional to its area \cite{Bekenstein-bhae, Hawking-particle-creation}. It tells us that there exists a microscopic description of the black hole with the number of the constituent microstates being the exponential of the entropy. 
\beq
S_{BH} = \frac{A}{4} \label{BHentropy}
\eeq

A thorough understanding of black hole microstates' features is a fundamental question in itself with important implications for quantum gravity. In this regard, AdS-CFT \cite{Maldacena:1997re, Witten-ads-and-holography, Gubser:1998bc} has provided us powerful tools to decipher features of black holes and quantum gravity in AdS in terms of boundary non gravitational observables.

Black hole information paradoxes \cite{Hawking:1974sw, Mathur:2009hf, Mathur:2012np} have traditionally served as beacons in the dark regarding physicists' quest for understanding quantum aspects of black holes, as well as for gravity in general. Every information paradox arises due to the existence of the black hole interior. Some important works discussing the black hole interior are \cite{Susskind:1993if, Susskind:1993mu, Page:1993wv, Strominger:1996sh, VanRaamsdonk:2010pw, Hayden:2007cs, Almheiri:2013hfa, Verlinde:2012cy, Shenker:2013pqa,  Maldacena:2013xja, Jafferis:2017tiu, Penington:2019npb, Almheiri:2019hni, Almheiri:2019qdq, Penington:2019kki,  Papadodimas:2012aq, Papadodimas:2013jku, Papadodimas:2013wnh, Papadodimas:2015xma, Papadodimas:2015jra}. In our work, we will address a close cousin of the information paradox colloquially known as the "Bags of gold" paradox \cite{Wheeler}, which serves as a valuable frame of reference illustrating the interwoven web of mysteries regarding the black hole interior.

We will briefly discuss the paradox now. Specific spacelike slices which go inside the black hole interior become very large in volume for a choice of boundary time.  Therefore these slices can host a considerable number of semiclassical excitations far higher than what the Bekenstein-Hawking entropy suggests, which leads to the paradox. We select these excitations such that they live far apart from each other on the Cauchy slices, thereby having zero spatial overlaps. The central question raised by the paradox is: How do we understand these states in the interior given that they are seemingly not reflected in the Bekenstein-Hawking entropy?

\begin{figure}

\centering
\includegraphics[width=.42\textwidth]{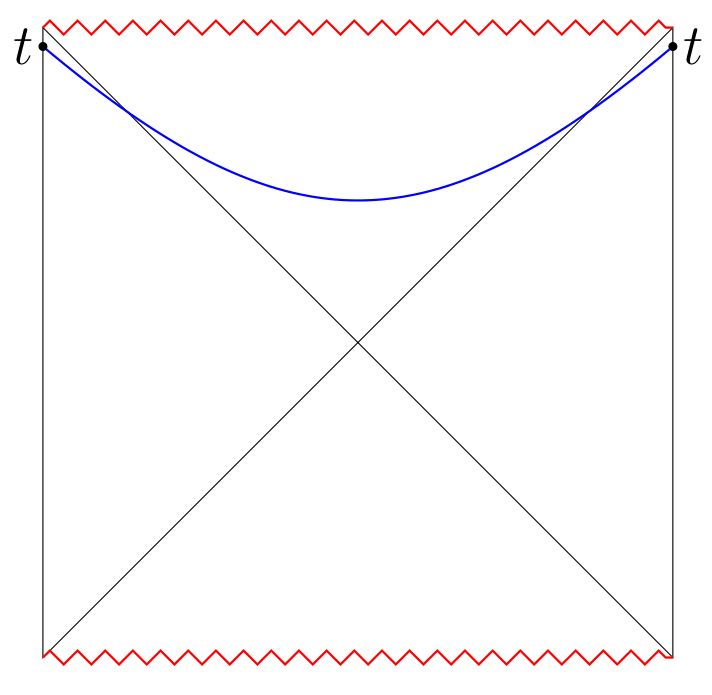}\hfill
\includegraphics[width=.42\textwidth]{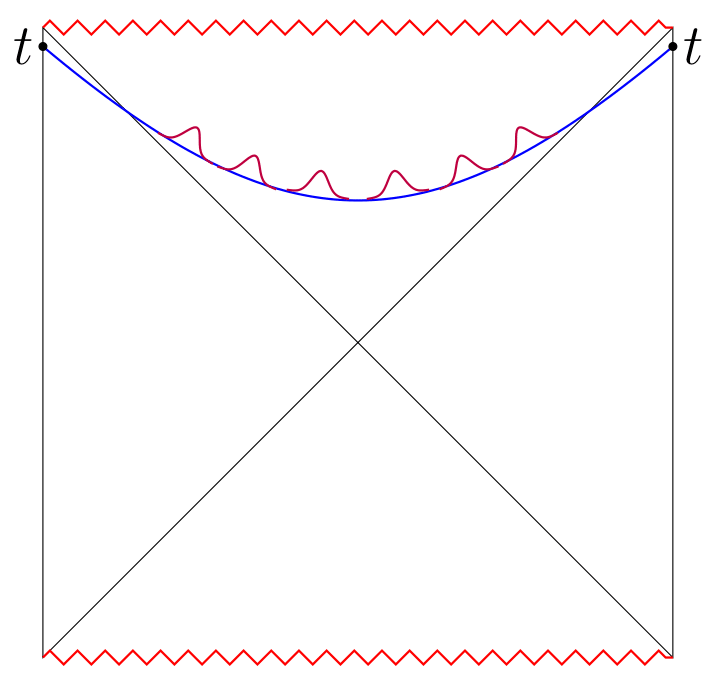}\hfill

\caption{The left figure displays the maximal volume slices for the eternal AdS Schwarzchild black hole. The volumes of these slices increasingly grow with boundary time $t$ thereby becoming very large at late times. The right figure demonstrates the bags of gold paradox for the eternal black hole on the maximal volume slices. We can accommodate an increasingly vast number of excitations placed far apart from each other on these slices which leads to the paradox.}
\label{paradoxfig}

\end{figure}

In this paper, we work with black holes in AdS where we formulate the paradox on "nice spacelike slices" of AdS black holes. These slices stay away from singularities and significant curvature invariants everywhere. We pose the bags of gold problem in this spacetime which allows us to utilize the AdS-CFT machinery to dissect the problem. We consider the eternal black hole \cite{Maldacena:2001kr} first, where we will demonstrate the paradox to its greatest extent by considering slices which possess the largest volumes for a given boundary.  Maximizing the spacelike volume for a given value of the boundary time constructs the aforesaid nice slice. The salient feature of such a slice is that its volume in the interior becomes increasingly large as the boundary time grows. Consequently, at late times we have slices with gigantic volumes. On these late time slices, we will fit in a high number of semiclassical bulk excitations placed spatially far apart from each other such that they have zero spatial overlap, and consequently are independent of each other. The number of such excitations is much more extensive than what is stated by the Bekenstein-Hawking entropy, which leads to our paradox.  Figure \ref{paradoxfig} displays the physical picture of the paradox. We are thus led to the question: Given that the Bekenstein-Hawking entropy is the area divided by 4, how do we account for the ever-increasing number of bulk excitations? Stated differently, does the entropy in equation \eqref{BHentropy} correctly count all these excitations or not?

In addition to the standard formulation of the bags of gold paradox as described above, we also argue that the effective field-theoretic description of the semiclassical excitations is inconsistent with the late time description of black holes using random matrix ensembles \cite{Bohigas, mehta2004random, doi:10.1063/1.1703773, doi:10.1063/1.1703775, MEHTA1960395, Hayden:2007cs, Shenker:2013pqa, Sekino:2008he, Lashkari:2011yi, Maldacena:2015waa, Cotler:2016fpe}. We will study spectral observables such as the energy level spacing distribution and the spectral form factor, which we expect to behave in specific fashions for Gaussian unitary ensembles  \cite{Bohigas, PhysRevLett.75.902, PhysRevE.55.4067, PhysRevE.56.264}. We will show that an EFT description of bags of gold excitations will violate these observables' expected features either qualitatively or quantitatively or in both fashions, thus leading to inconsistencies.

Our proposed resolution to the above paradoxes is that we have tremendously overcounted the bulk states in the interior.  Semiclassical bulk states placed far apart from each other in the interior are seemingly orthogonal. However, these states have small and significant inner products between them, which deviates from the semiclassical expectation of zero inner products. This is because in gravity, two coherent states corresponding to even vastly different classical configurations have a small non-vanishing inner product. In other interactions such as electrodynamics, two such coherent states can have a vanishingly small inner product.  In contrast, the inner product between coherent states in gravity does not go to zero but saturates to an O$(e^{-\frac{S}{2}})$ number. The non-vanishing of inner products between two sufficiently distinct coherent states is the primary reason leading to overcounting.  More generally, we will show that the maximum number of vectors with small inner products that can be accommodated in a Hilbert space is exponentially larger than the dimension of the Hilbert space. This kinematical statement justifies the existence of an enormous number of interior bulk excitations leading to our paradox. As an example, if the bulk Hilbert space's true dimensionality is $e^S$ and the inner products between bulk excitations are of order $e^{-\frac{S}{4}}$, then the maximum number of bulk excitations $(m)$ with such small inner products is a vast number given by \footnote{  \S \ref{resolution} gives the details of this calculation.}:
\beq
 m \sim  e^S \times \exp{\frac{e^{\frac{S}{2}}}{2}}.
 \eeq
 
 If we consider even a small system with dimension $e^S = 3.6 \times 10^5$ with inner products of the order $e^{-\frac{S}{4}}$ then we can fit in up to $10^5 \times e^{300}$ vectors in the Hilbert space which is a huge number, far more sizeable than the number of atoms in our known observable universe ($\sim 10^{78}$ - $10^{82}$).

 We will pause here to discuss some points. A natural extension to our present discussion is to study the paradox for single-sided black holes. We will do so using a setting similar to the eternal black holes, which we display in Figure \ref{pureparadox}. We advocate the same resolution for the single-sided paradox as we have for the eternal case. Another point is that there exists an entirely different way to arrive at this paradox. The paradox also arises if we glue inflating or FLRW regions inside the interior by using junction conditions \cite{Marolf-bog, Hsu:2009kv, Freivogel:2005qh, Fu:2019oyc,  Ong:2013mba}. These glueings result in similar spacelike slices which have huge volumes in the interior. As a consequence of the paradox, it is also argued that the CFT does not contain the interior states. In our work, we assume that a state-dependent map reconstructs the black hole interior, thus describing the states behind the horizon \cite{Papadodimas:2012aq, Papadodimas:2013jku, Papadodimas:2013wnh, Papadodimas:2015xma, Papadodimas:2015jra}. Thus our interior does not have any glued regions, and the CFT captures our interior excitations. On a related note, \cite{Langhoff:2020jqa, Nomura:2020ewg} also discuss various subtleties regarding the problem of large interior volumes and advocate a similar resolution.

\subsection*{Overview of results}
We will now give a quick overview of our results. \S \ref{maxvolume} poses the paradox discussed above for eternal black holes in detail. \S \ref{resolution} discusses our proposed resolution, where we also determine the maximum number of vectors that can be fit inside a Hilbert space with small inner products. In \S \ref{CFT}, we show that the paradox does not show up in the fine-grained entropy of the CFT. From the CFT perspective, the action of state-dependent operators on the state of the black hole generates the interior bulk states in our construction. We show that the bulk state produced by the action of interior operators on the thermofield double state \cite{Maldacena:2001kr, Israel:1976ur, Takahasi:1974zn} does not lead to any change in the Von Neumann entropy of the CFT. We also calculate the fine-grained entropy using quantum extremal surfaces for the eternal black hole. These surfaces do not enter the black hole interior and therefore, do not capture our interior excitations. Consequently, there is no paradox in the dual CFT.  These observations strongly support our claim that the interior states arise due to overcounting and are not independent excitations in quantum gravity.

In the bulk description, it is essential to understand the behaviour of the interior excitations. From the CFT perspective, our excitations appear to be in equilibrium when probed using simple operators in the right side CFT. However, they are out of equilibrium when probed with operators belonging to the complement of the \textit{small algebra} of \textit{simple operators} \footnote{See Appendix \ref{StateDepReview} for the definition of simple operators, the small algebra and its complement}. These properties of the excitations lead to a bulk picture of the excitations arising from the left past horizon of the eternal black hole and travelling through the left side.  Afterwards, they fall into the left future horizon where they go on and intersect the nice slices. The above nature of the excitations physically demonstrates the paradox in Figure \ref{eternalparadox} where different excitations come out of the left horizon at particular times governed by the unitary operator $U(t)$. The initial bulk state of the excitations on the black hole is a Euclidean black hole glued to the Lorentzian geometry \cite{Hartle:1976tp, Hartle:1983ai}. Here the excitations are generated using operators at the Euclidean AdS boundary (See Figure \ref{hh}).

We estimate that two excitations placed far apart on the nice slices of single-sided black holes have an overlap of O$(e^{-\frac{S}{2}})$. Such an overlap strongly backs our resolution involving small inner products and is the topic of \S \ref{singleside}. We discuss how the treatment of bags of gold excitations using effective field theory violates black holes' expected spectral properties in \S \ref{spectralsff}. We provide some toy examples of bags of gold configurations there, which violate the qualitative and quantitative features of spectral form factor and energy level spacing distribution. We also argue how our resolution fixes these issues. Next, we explicitly demonstrate that there can be a large number of excitations living in the black hole interior using toy models in \S \ref{matrixmodel}. These toy models are small $N$ matrix models in which we first construct a typical state \cite{lloyd2013pure, PhysRevLett.54.1350, PhysRevLett.80.1373} in order to model single-sided black holes. We then use the typical state and the small algebra to construct the \textit{small Hilbert space} describing interior bulk excitations. Random combinations of operators living on this small Hilbert space gives rise to smeared bulk excitations. We see that the small Hilbert space can embed a large number of states having small inner products with each other. We then construct states resembling excitations placed far apart from each other on the Cauchy slice in these matrix models. These states have small inner products, thereby confirming our resolution discussed in \S \ref{singleside}.

It is a natural question to ask why such an overcounting does not occur for quantum statistical systems and is special to black holes. Consider a statistical system which has a Hilbert space of dimension $m$. One can apply our resolution to this system and ask whether this system has a much smaller dimension $n$, with $m \gg n$. While we can kinematically pose such a statement, such a situation leads to discrepancies in thermodynamic observables. Another consequence of such a modelling is that forbidden quantum state transfers can occur in the larger system modelled with $n$ vectors. We discuss these issues in \S \ref{genproperties}.

\section{The bags of gold paradox for the eternal black hole}
\label{maxvolume}

In this section, we will outline the construction of the maximal volume surfaces. Afterwards, we will place excitations on these slices. Lastly, we will pose and discuss the paradox in detail.

\subsection{Maximum volume slices in the interior}
Consider an eternal black hole at boundary time $t$ as in Figure \ref{paradoxfig}. We want to construct nice slices which stay away from singularity everywhere and possess the maximum volume for a given boundary time $t$. We will work with the AdS Schwarzchild metric in $d+1$ dimensions is given by
\begin{equation}
ds^2 = -\frac{4f(r)}{f'(r_h)^2} \, e^{-f'(r_h) r^*} du_k \, dv_k + r^2 d\Omega^2_{d-1} , 
\end{equation}
where $f(r) = r^2 + 1 - \frac{C}{r^{d-2}}$, $r_h$ is the black hole horizon and $r^*$ is the tortoise coordinate. The subscript $k$ denotes Kruskal coordinates. Our goal is to show is that the interior's volume grows as we increase the boundary time $t$.

Since the paradox involves only the interior, a demonstration of the growth of the interior volume will be sufficient for our purposes. Instead of parametrizing the slices with the boundary time, we will parametrize them using the Kruskal coordinates $(u_0,0)$ on the left horizon and $(0,v_0)$ on the right horizon, as shown in Figure \ref{eternalparadox}. Thus we change our problem to a similar one where we compute the maximum volume of slices which end at $(u_0,0)$ on the left horizon and $(0,v_0)$ on the right horizon. This problem has two advantages. We see the first advantage of calculating the maximal volume surfaces in the case of single-sided black holes in \S \ref{singleside}. These black holes possess the entire interior region but do not have a boundary time on the left. Therefore we can utilize this construction of maximum volume slices for the single-sided case. This problem also overcomes the problem of infinite exterior volumes \footnote{though this can also be tamed by introducing a boundary cutoff.}.

We set $u_0 =v_0$ using the isometry of AdS spacetime. It is convenient to use the infalling Eddington-Finkelstein coordinate $v = t+ r^*$ in order to calculate the maximum volume surfaces.
\begin{equation}
ds^2 = -f(r)\, dv^2 + 2 dr\, dv + r^2 d\Omega^2_{d-1}
\end{equation}

Note that $v$ here is different from the Kruskal coordinate $v_k$. We define an affine spacelike parameter $\sigma$ to parametrize the nice slice. We now need to extremize the following volume integral to obtain the maximum volume of these surfaces.
\begin{equation}
V = V_{d-1}\int d\sigma \, r^{d-1} \left( -f(r) \, \dot{v}^2 + 2\dot{r} \, \dot{v}\right)^{\frac{1}{2}},  \label{integrandlagrangian}
\end{equation}
where $V_{d-1}$ is the volume of the $(d-1)$ spherical ball. We end up with the following expression for the volume \cite{Stanford:2014jda, Carmi:2017jqz, Susskind:2018pmk}:
\begin{equation}
V = \frac{\beta A(r_{min})}{2 \pi}  \log{u_0} + \text{O}\,(1) \label{e1}
\end{equation}
where $A(r_{min})$ and $\text{O}\,(1)$ are terms of order one which do not grow with $u_0$. In equation \eqref{e1}, $r_{min}$ is determined using 
 \begin{equation}
 f(r_{min}) \, r^{2(d-1)}_{min} + E^2 =0,
 \end{equation}
where $E$ is a conserved quantity $E = -\frac{\partial L}{\partial \dot{v}}$ with $L$ denoting the integrand of equation \ref{integrandlagrangian}. The volume extremization, derivation of the resulting equation \eqref{e1} and $A(r_{min})$ are calculated in Appendix \ref{Appendix 1}. The important observation here is that the interior volume of the nice slice increasingly grows with the Kruskal time. The physical reason is that the wormhole grows larger and larger with Kruskal time.

\subsection{Placing semiclassical excitations on the nice slice}
\label{excitations}

Since the volume of the nice slice in the interior keeps increasing with the Kruskal time, the interior can accommodate an increasingly large number of semiclassical excitations far apart from each other such that their spatial overlaps with each other are zero. At late times the slice's volume goes to infinity, and therefore a high number of excitations can be placed far apart from each other. These interior excitations are created by acting with unitary operators on the right CFT in the thermofield double state. Eqn. \eqref{ex1} describes an interior excitation,
\beq
\ket{\psi^e_1} = C_1 \, e^{-\frac{\beta H_R}{2}  }\,  U_1(t_1) \, e^{\frac{\beta H_R}{2}  } \ket{\text{TFD}} \label{ex1}.
\eeq

Here $U(t_1)$ is an unitary operator acting on the right CFT, $C_1$ is the normalization constant, $H_R$ is the right CFT's Hamiltonian. The state $\ket{\psi_e}$ represents our excitation. These states are motivated by the state-dependent formalism, which we review in Appendix \ref{StateDepReview}. The unitary operator $U_1$ controls the position of these excitations on the slice. This control results due to the timelike coordinate $t$ in the exterior becoming a spacelike coordinate in the interior. We now create another excitation in the interior:
\beq
\ket{\psi^e_2} = C_2 \, e^{i H_R T} \, e^{-\frac{\beta H_R}{2}  }\,  U_2(t_2) \, e^{\frac{\beta H_R}{2}  } \, e^{-i H_R T} \ket{\text{TFD}}.
\eeq

The action of the Rindler Hamiltonian using factors of $e^{-i H_R T}$ spatially separates this second excitation from the first one. Since the exterior timelike coordinate becomes spacelike in the interior, these excitations are placed far apart from each other if $T$ is large enough. We now generate $m$ number of such excitations similarly, with each excitation placed far apart from the previous one as a result of modulating with the factor $e^{-iHT}$, where $T$ denotes the time difference between consequent excitations. We will discuss the nature of these excitations in more detail in \S \ref{initialHH}.  Therefore a physical picture of placing the excitations on the nice slice is as follows: Consider an excitation created at time $t_1$. This excitation proceeds into the interior of the black hole and intersects the nice slice. Other excitations are created at time $t_2$, and so on at $t_3, t_4$ up to $t_m$. 

 \subsubsection{Physical properties of the excitations}

 We now demand certain physical properties which these excitations should satisfy. We generate the excitations such that the backreaction is very small as compared to the mass of the black hole. If we have $m$ excitations each having energy of the order of $E_0$, then the condition for preventing backreaction is given by
 \beq
 m \, E_0 \ll M_{BH}.
 \eeq

 We will ensure that the density of excitations $\rho = \frac{m}{V}$ is a finite quantity in the thermodynamic limit, i.e. with $m$ and $V$ large. Fixing the density allows us to calculate the entropy of these excitations in the effective field theory approximation by treating the system as a "dilute gas" of excitations living on the nice slice of the black hole. We also want that the separation between any two excitations is quite more substantial than the smearing time scale $\delta t$ associated with each excitation which leads to the following condition. 
\beq
\delta t_i \ll |t_i - t_j|, \quad \forall j \neq i \label{timesep}
\eeq

 We also impose an IR cutoff for the excitations which restricts them completely to the interior of the black hole. In the late time limit, we demand that the excitations have a length scale shorter than the volume of the black hole divided by the number of excitations, which gives rise to the following bound:
\begin{equation}
\frac{V}{m \, V_{d-1}} \gg \frac{1}{E_0}.
\end{equation}
where $V_{d-1}$ is the volume of the unit spherical ball as defined previously. Thus our construction defines a "dilute gas" of excitations living in the black hole interior, such that each of these excitations has zero spatial overlap with the others. We will clarify further details regarding the physical behaviour of the excitations in the bulk in \S \ref{initialHH}.

\subsection{The paradox in the bulk}
\label{paradox}
We will now roughly calculate the entropy of the "dilute gas" of excitations in the bulk interior using the microcanonical ensemble, assuming that our excitations behave classically. Let $ E$ denote the total energy of the configuration. The volume $\Sigma_p$ of a shell with uncertainty $\Delta_E$ centred about $E$ in the momentum space is given by:
 \begin{equation}
 \Sigma_p = \frac{\sqrt{m}}{(m-1)!}\, E^{m-1} \, \Delta_E.
 \end{equation}

Using this we calculate the volume of the phase space spanned by the gas.
\begin{equation}
\Omega(E,V,m) = \frac{V^m}{m!} \frac{\sqrt{m}}{(m-1)!}\,  E^{m-1} \, \Delta_E.
\end{equation}

The phase space volume enables us to calculate the entropy of the ensemble. We use Stirling approximation and ignore the subleading terms in $m$. Finally, the expression for entropy with $\rho = \frac{m}{V}$ is obtained to be
\begin{equation}
S(E,V,m) = m \log{\frac{V \, E}{m^2}} = m \log{\frac{V \, E_0}{m}} = V \rho \log{\frac{E_0}{\rho}}, \label{paradoxinentropy}
\end{equation}
where $E = m E_0$, with $E_0$ being the average energy of a single excitation. Since we have imposed that the density $\rho$ is a finite non-zero quantity, \eqref{paradoxinentropy} indicates that the entropy scales as the volume. This scaling gives rise to the paradox that the entropy of the dilute gas is larger than the Bekenstein Hawking entropy of the black hole at late times.

\section{The resolution: Overestimation of the Hilbert space's dimensionality}
\label{resolution}

The reason why the paradox arises is due to a colossal overcounting of the bulk Hilbert space. In our construction, we ensured that the semiclassical excitations have zero spatial overlap, which is sufficient for two different excitations to be independent in effective field theory, i.e. with a vanishingly small inner product. This section motivates why this assertion is not correct in quantum gravity and demonstrates that we can embed many more vectors in a Hilbert space with small inner products than given by the dimension of the space. Some results in this section were also discussed in unpublished notes in \cite{CR}.

We first review why semiclassical gravity predicts that the inner product between two vectors in the Hilbert space can be arbitrarily small. Afterwards, we will look at why such a prediction does not hold true in quantum gravity.

\subsubsection*{Inner products in semiclassical gravity}

We will follow the work of \cite{Papadodimas:2015jra} here in order to compute the inner product between semiclassical states. We work with a background metric $g^0_{\mu \nu} (x)$ in $d+1$ dimensions, and consider small linearized fluctuations $g^{'}_{\mu\nu} = g^0_{\mu \nu} + \sqrt{8\pi G_N} \delta g_{\mu \nu}$ about it. In general these linearized fluctuations can be expressed in terms of creation and annihilation operators
\beq
\delta g_{\mu \nu}(x) = \sum_i \sum_k a^i (k)\,  g^{i}_{\mu \nu} (k) + \text{h.c.}
\eeq where $i$ denotes the $\frac{(d+1)(d-2)}{2}$ polarizations and $k$ goes over the momenta. We choose the functions $g^{i}_{\mu \nu} (k) $ such that the creation and annihilation operators obey the same commutation relations for a simple harmonic oscillator. We will look at the coherent states formed by the action of the creation operators which creates the excited spacetime:
\beq
\ket{\alpha} = C_{\alpha} \, e^{\, \sum_i \sum_k {a_i}^{\dagger} (k) \, \alpha (k)} \ket{0},
\eeq
 where $C_{\alpha}$ is the normalization constant and $\ket{0}$ is the vacuum such that ${a_i} (k) \ket{0} = 0$. The expectation value of the metric operator on a coherent state $\ket{g^{\text{cl}}}$ gives us the classical value of the metric:
 \beq
 g^{\text{cl}}_{\mu \nu} = \braket{g^{\text{cl}}\, |\hat{g}_{\mu \nu}(x)|\, g^{\text{cl}}}.
 \eeq
 
  We now consider the inner product between the background spacetime and the excited spacetime, such that the two spacetimes are "distant" in the phase space. Here "distant" means a substantial classical perturbation $\delta g_{\mu \nu} \sim \frac{\Delta}{\sqrt{8\pi G_N}} = \Delta \, \mathcal{N}$, where $\mathcal{N}$ is the central charge of the CFT ($\mathcal{N} = N^2$ for gauge theories with gauge group $N$). For small linearized fluctuations, we set $\Delta \sim \text{O} \, (1) \ll \text{O} (\mathcal{N})$ such that $\Delta \ll 1$, which still allows us to do linearized perturbations while not being vanishingly small. As shown in \cite{Papadodimas:2015jra}, the semiclassical inner product between the two bulk states is given by
 \beq
 \braket{g^{0}_{\mu \nu} \, |g^{\text{cl}}_{\mu \nu}} = \exp{-\mathcal{N} v\left(g^0, \, g^{cl}  \right)}
 \eeq
 where $v\left(g^0, \, g^{cl} \right)$ is an $\text{O} \, (1) $ quantity. Thus we conclude that the inner product between two different semiclassical excitations can be arbitrarily small. This is a feature common to a QFT, coherent states corresponding to quite different classical excitations can have a vanishingly small overlap.
 
 \subsubsection*{Inner products in quantum gravity from the CFT description}
 
 Using the dual CFT description, we will see why the analysis in the preceding subsection is misleading when the phase space "distance" between the classical configurations becomes large. Contrary to the semiclassical indication, the inner product between two different vectors might be a small but finite number even if the classical description is completely different \cite{motl, Papadodimas:2015xma, Papadodimas:2015jra}. A simple example is the overlap between two factorized AdS spacetimes and the thermofield double, which are very different classical configurations. These two have an overlap given by:
 \beq
 \braket{0,0\, |\text{\text{TFD}}} = \frac{1}{\sqrt{Z(\beta)}},
 \eeq
 which is small but nonvanishing. The physical basis behind this small overlap is the following: the semiclassical inner product is obeyed only up to a particular "distance" in the phase space between two different classical configurations. Beyond this distance, inner products are saturated and differ from the semi classical inner product.

 An example of this saturation is given by "time-shifted states" in the CFT \cite{Papadodimas:2015xma, Papadodimas:2015jra}, which represent different bulk configurations. Consider the time shifted state given by time evolution on the left CFT acting on the thermofield double: 
 \beq
 \ket{\psi_T} = e^{iH_L T} \ket{\text{TFD}}.
 \eeq
 
 On the thermofield double consider $m =e^S$ distinct time shifted states each shifted by a time $ \left(\, T_1, T_2 \, ... \, T_m \right)$. Now there exist a solution for $\alpha_i$'s given in the following equation:
 \beq
 \abs{\ket{\text{TFD}} - \sum_{i =1}^{e^S} \alpha_i \, e^{iH_L T_i} \,  \ket{\text{TFD}}}^2 = \text{O} \, (e^{-\mathcal{N}}).
 \eeq
 
 This leads to the inner products developing a saturated "fat tail" of magnitude $\text{O} \, (e^{-\mathcal{N}})$ which is our primary motivation for overcounting. This shows that these bulk states are not really independent of each other.

 We give another proof of the presence of small inner products from the CFT description in \S \ref{typicalovercounting} for single-sided black holes. Given a CFT dual to a single-sided black hole, we will show that two far apart excitations have an inner product of the order of O$(e^{-\frac{S}{2}})$, which serves as the basis for overcounting in the single-sided black holes.

 The fundamental reason why this saturation of inner products happens in gravity is an obstruction to the lifting of classical observables living on the phase space to the Hilbert space. The $d$-metric and its canonical conjugate momentum in the $d+1$ ADM decomposition cannot be naively lifted to well-defined operators on the Hilbert space, as they give rise to the semiclassical inner product. Apart from these examples, there also exist other cases where the inner product in effective field theory receives small corrections in quantum gravity.  This "fat tail" is similar to the "spectral form factor" in \cite{Cotler:2016fpe}. Another striking example is the statement that two states in quantum gravity might turn out to be the same \cite{Itzhaki:2019cgg}. 
 
 \subsection{How many bulk excitations can we possibly have?}
 
  We saw in the preceding subsection that all distinct bulk excitations are not independent of each other.  Since the inner products saturate, taking excitations far apart would not make them independent. With this motivation, it becomes a natural question to ask how many bulk excitations can we fit inside a Hilbert space of dimension $n$.

 This question has a profound consequence: a black hole with coarse-grained entropy $S_{BH}$ can still have a vast number of bulk excitations living on the nice slices, and hence there is no paradox. 

\subsubsection{How many vectors can we fit inside a Hilbert space of dimension $n$?}
\label{howmanyexcitations}

We consider the following problem: In a Hilbert space $\mathbf{H}$ of dimension $n$, what is the maximum number $m_n(\epsilon) \equiv m$ of vectors $\{v_i\}$ which satisfy the following relations: 
\begin{equation}
\braket{v_i|v_i} =1 \quad \& \quad \abs{\braket{v_i|v_j}} \leq \epsilon, \quad i \neq j. \label{innpro}
\end{equation}

We have $m_n(0) = n$ trivially. The solution to this problem is as follows. Unit vectors in the Hilbert space live on the surface of an $(2n-1)$ dimensional real sphere. We can fix one vector to be $\ket{v_1} = (1,0, \dots ,0,0)$. The remaining vectors $\ket{v_i} = \lc a_1, a_2, \, \dots \,, a_{n-1}, a_n \rc$ will satisfy the following equation,
\begin{equation}
\abs{a_1}^2 + \abs{a_2}^2 + \, \dots \, + \abs{a_{n-1}}^2 + \abs{a_n}^2 = 1.
\end{equation}

For $i \neq 1$, \eqref{innpro} implies that $|a_1|^2 \leq \epsilon^2$. Therefore around a vector $\ket{v}$, there is an exclusion zone where there can be no other vector. The boundary of this region is given by
\begin{equation}
\abs{a_2}^2 + \, \dots \, + \abs{a_{n-1}}^2 + \abs{a_n}^2 = 1 - \epsilon^2.
\end{equation}

Since $a_k \in \mathbb{C}$, we write $a_k = c_k + id_k$, where $c_k, \, d_k \in \mathbb{R}$. We perform the worst-case estimate of the number of vectors by assuming all the inner products are of the order $\epsilon$. We obtain the naive estimate for the number of vectors that satisfy the inner product bounds by dividing the surface area of the $2n-1$ dimensional real sphere (since $\abs{a_1}^2 \sim \epsilon^2$) with the area of the exclusion zone. The exclusion zone for each vector has the radius $\frac{\epsilon}{2}$. Therefore each sphere will have the volume given by
\begin{equation}
\sum_{k = 2}^{n} c_k^2 + d_k^2 = 1 - \frac{\epsilon^2}{4}.
\end{equation}

A more accurate computation would also require the packing fraction of such exclusion zones. One can then count the number of vectors and multiply it by the packing ratio to approximately get the highest number of vectors.
 
 \begin{equation}
 m \approx  \frac{P \, S_{2n-1}}{V_{2n-2}} \left(\frac{1}{1-\frac{\epsilon^2}{4}}\right)^{2n-2} = 2 \pi P \left( 1-\frac{\epsilon^2}{4}\right)^{-2n+2}
 \end{equation}
 
 Here $S_{2n-1}$ is the surface area of the $(2n-1)$ dimensional sphere, the volume enclosed by the $(2n-3)$ dimensional sphere is given by $V_{2n-2}$ and $P$ denotes the constant of proportionality which gets contribution from the packing fraction and also takes into account small errors which may have resulted from our rough counting method. We have also used $\frac{ S_{2n-1}}{V_{2n-2}} = 2\pi$. Let us have a look at the function $\left( 1-\frac{\epsilon^2}{4}\right)^{-2n+2}$. We are interested when $n$ becomes very large. Now using the definition of the exponential function we obtain
 \begin{equation}
 \lim_{n\to \infty} \left(\frac{1}{1-\frac{\epsilon^2}{4}}\right)^{2n-2} = \lim_{n\to \infty} \left(1-\frac{\frac{n\epsilon^2}{4}}{n}\right)^{-2n+2} \approx \lim_{n\to \infty} \left(1-\frac{\frac{n\epsilon^2}{4}}{n}\right)^{-2n} = e^{\frac{n\epsilon^2}{2}}.
 \end{equation}
 
 Note that the above expression is valid for any value of $\epsilon$, including our case where $n\epsilon^2 >>1$. We evaluate the value of $m$ in the limit of large $n$ to be
  \begin{equation}
 m \approx 2\pi P \,  e^{\frac{n\epsilon^2}{2}}. 
 \end{equation}
 
 Since our small inner products in question are very close to zero, i.e. $\epsilon \approx 0$, we fix the proportionality constant $2\pi P$ in the case when $\epsilon = 0$,  which sets $2\pi P = n$. Therefore the formula describing maximum possible vectors for small $\epsilon$ is given by
 \begin{equation}
 m \approx n \, e^{\frac{n\epsilon^2}{2}}. \label{ned}
 \end{equation}

 We pause here to reflect upon what our formula in \eqref{ned} tells us. With tiny inner products $\epsilon  \ll 1$ such that $\epsilon > e^{-\frac{S}{2}}$, we can obtain an extremely enormous overcounting of the Hilbert space. As an example consider  all  inner products $\epsilon \sim e^{-\frac{S}{4}}$. The maximum number of states with such a small inner product that can be embedded in the Hilbert space of dimensionality $e^S$ is given by:
 \beq
  m \sim  e^S \times \exp{\frac{e^{\frac{S}{2}}}{2}}.
 \eeq
 
 This counting suggests that even for a small $S$ like $S = 10^5$, $m$ is a vast number. Thus a high number of bulk states can be embedded in the actual smaller Hilbert space with tiny inner products, which is the surprising fact underlying our resolution.

 In low dimensions equation \eqref{ned} seems to contradict our intuition, for we do not see such tremendous growth. Appendix \ref{lowdim} deals with the calculation of inner products for vectors denoting the corners of regular polyhedra in general dimensions while building up from low dimensional examples. Inner products of these corner vectors of regular polyhedra eventually reproduce equation \eqref{ned} when the dimensionality becomes large. This approach helps develop our intuition for large Hilbert spaces since it builds up starting from low dimensional examples.

 \section{Resolution of the paradox from the boundary perspective}
 \label{CFT}
 
 As mentioned in the introduction, equation \eqref{BHentropy} is the coarse-grained entropy of a black hole. The origin of coarse-grained quantities like the thermodynamic entropy is due to inherent sloppiness since we measure only a small subspace of the Hilbert space. As a result, coarse-grained quantities can grow under unitary time evolution. In contrast, the fine-grained entropy or the Von Neumann entropy is a more accurate measure of the degrees of freedom. The fine-grained entropy remains invariant under unitary time evolution.

  We hereby digress to investigate the paradox from the boundary viewpoint and calculate the Von Neumann entropy on the CFT side. We will show that the calculation of the entropy of the CFT reveals the absence of any paradox because the insertion of the excitations on the thermofield double preserves the Von Neumann entropy.

  Computation of the generalized entanglement entropy also demonstrates that there is no paradox in the CFT. This computation involves a choice of quantum extremal surfaces and does not depend on the precise details of the excitations.

  We note an important point here: The proof that there is no paradox in the boundary does not capture the qualitative picture of the paradox in bulk. However, this indicates a crucial fact: the excitations do not increase the fine-grained entropy. The invariance of fine-grained and coarse-grained entropy along with the assumption that state-dependent operators reconstruct the black hole interior leaves us with no choice apart from overcounting of vectors to resolve this paradox.

 \subsection*{CFT excitations: No paradox}
 
 In this subsection, we will look at the entanglement entropy of the right CFT. Consider the thermofield double state, which consists of the left and the right CFTs. Tracing over the left region gives us the reduced density matrix for the right CFT, which is the thermal density matrix $\rho_T$.
\beq
\rho_R = \text{Tr}_L \ket{\text{TFD}}\bra{\text{TFD}} = \sum_i \frac{e^{-\beta E_i}}{Z(\beta)} \ket{E_i}_R\bra{E_i}_R = \rho_T, 
\eeq
 where $\rho_T$ is the thermal density matrix. Equation \eqref{ex1} describes an excitation in the interior:
 \beq
\ket{\psi^e_1} = N_1 \, e^{-\frac{\beta H_R}{2}  }\,  U^1_R \, e^{\frac{\beta H_R}{2}  } \ket{\text{TFD}},
\eeq

We will define the following unitary operators for our convenience:

\beq
V_R^i \equiv e^{i H_R T} U_R^i e^{-i H_R T}.
\eeq

Note that here we have included the time evolution contributions $e^{-iH_R T}$'s inside the unitary $V$'s since they represent unitary contributions. Till now we have worked in the semiclassical picture where we have treated the excitations as $m$ different vectors. However from the CFT perspective, the boundary state with $m$ interior excitations is written as the action of a single interior operator on the thermofield double state. The following expression is due to the specific form of the interior operators:
\beq
\ket{\psi^e} =  C \, e^{-\frac{\beta H_R}{2}  }\,  V^{\, m }_R (\,t_m) \, V^{\,m-1}_R (\,t_{m-1}) \, \dots \, V^{\,2}_R(t_2) \, V^{\,1}_R (t_1)\, e^{\frac{\beta H_R}{2}  }  \ket{\text{TFD}} =   C \, e^{-\frac{\beta H_R}{2}  }\,  V_R \, e^{\frac{\beta H_R}{2}  } \ket{\text{TFD}} \label{manyexcitations}
\eeq
We now calculate the reduced density matrix on the right region for this system of excitations. 
\beq
\begin{split}
\rho'_R = \text{Tr}_L \ket{\psi^e}\bra{\psi^e} &=  \abs{C}^2 \, e^{-\frac{\beta H_R}{2}  }\,  V_R \, e^{\frac{\beta H_R}{2}  } \,\text{Tr}_L \ket{\text{TFD}}\bra{\text{TFD}} \, e^{\frac{\beta H_R}{2}  }\,  V^{\dagger}_R \, e^{-\frac{\beta H_R}{2}  }\\
&= \frac{1}{Z(\beta)} \, e^{-\frac{\beta H_R}{2}  }\,  V_R \, e^{\frac{\beta H_R}{2}  } \,e^{-\beta H_R} \, e^{\frac{\beta H_R}{2}  }\,  V^{\dagger}_R \, e^{-\frac{\beta H_R}{2}  }\\
&= \rho_T.\\
\end{split}
\eeq

 The above manipulations follow because $V_R$ is a unitary operator. We expect the thermal density matrix to remain unchanged under interior operator insertions because the thermal behaviour arises due to the horizon's existence and is irrespective of insertions in the interior unless a large backreaction changes the horizon. The interior operators are defined only in the effective field theory limit, i.e. the backreaction is small, and hence the thermal density matrix remains invariant. Since the density matrix itself does not change due to the excitations, the entanglement entropy does not change as well. Therefore we see that there is no paradox in the CFT as interior excitations do not change the entanglement entropy.

\subsection*{Generalized entanglement entropy of the CFT}

  Using the generalized entanglement entropy \cite{Ryu:2006bv, Ryu:2006ef, Hubeny:2007xt, Lewkowycz:2013nqa, Faulkner:2013ana, Engelhardt:2014gca}, we can again show that there is no paradox in the CFT. Quantum extremal surfaces are defined as surfaces which extremize the sum of area and bulk entanglement entropy contributions, given a boundary subregion $B$. This extremized sum is the generalized entanglement entropy of $B$.

\beq
S_{\text{gen}}(B) = \text{Min}_X \, \text{Ext}_X \, \ls \frac{\text{Area}(X)}{4 G_N} + S_{\text{bulk}}\lc \Sigma_B^X\rc\rs
\eeq

Consider $B$ to the right boundary region $R$ on which the right CFT lives. We will consider the case with no excitations living on the black hole first. Quantum extremal surfaces for this case end at the horizon, therefore the generalized entanglement entropy is given by:
\beq
S_{\text{gen}}( R) = \frac{\text{Area of black hole}}{4 G_N} \label{fge}
\eeq

 Now consider a situation where the matter content due to excitations in the interior is very large, which is our case of interest. Consequently, the bulk entropy $S_{\text{bulk}}$ in the interior of the black hole due to all the excitations is very large.  In this case, the quantum extremal surfaces are no different and go only up to the horizon, thereby not capturing $S_{bulk}$ in the interior region. As a result, the fine-grained entropy again is given by \eqref{fge}. Therefore we conclude that there is no bag of gold paradox. We note that we do not need the precise form of the excitations in order to derive this conclusion.

\section{The nature of the excitations and the initial bulk wavefunction}
\label{initialHH}

\begin{figure}

\centering
\includegraphics[width=.8\textwidth]{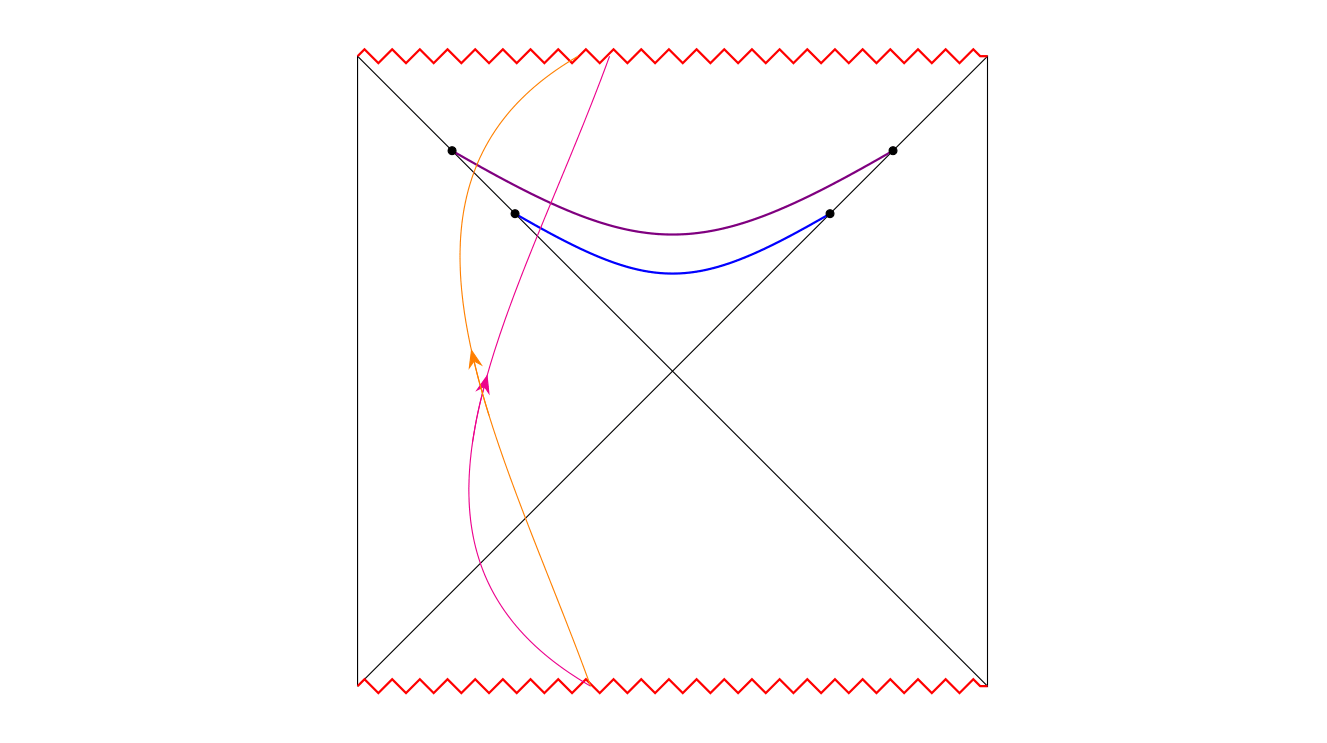}\hfill

\caption{Bulk excitations denoted by orange and magenta lines arise from the left past horizon and fall into the left future horizon. These states come out of equilibrium as indicated in equation \eqref{ooe} around the time $t_1$ and $t_2$ for the orange and magenta excitations. In the bulk this out of equilibrium behaviour is indicated by how far the excitations protrude out on the left. The unitaries control the position of the excitation on the slice, and large $\abs{t_1-t_2}$ leads to large spatial separation. Interior excitations at late times are visible only to later slices. Consequently, we can keep accomodating more and more excitations at later and later times which leads to the paradox.}
\label{eternalparadox}

\end{figure}

 In this section, we are interested in understanding the exact nature of the excitations created by interior operators as given in \eqref{ex1}. The excitations' behaviour also holds the key to qualitatively understand the initial state in the bulk, which leads to the paradox. From the CFT perspective, the states given in \eqref{ex1} are non-equilibrium states \cite{Papadodimas:2017qit}, which we briefly describe. These states $\ket{\psi^e_1} = C_1 \, e^{-\frac{\beta H_R}{2}  }\,  U_1(t_1) \, e^{\frac{\beta H_R}{2}  } \ket{\text{TFD}}$ arise from the past left horizon and end up at the future left horizon. To see this, we first show that these excitations are invisible to the small algebra $\mathcal{A}_R$ \footnote{The algebras $\mathcal{A}$, $\mathcal{A'}$ and the associated small Hilbert space formed by acting with them on the thermofield double state are reviewed in Appendix \ref{StateDepReview}.}. Therefore the time dependence of these observables cannot be seen by probing with $O \in \mathcal{A}_R$.

\begin{equation}
\frac{d}{dt}\braket{\psi^e_1|\, O(t)\,|\psi^e_1} \sim \text{O}\left(\frac{1}{S}\right)
\end{equation}

However, these states are truly non-equilibrium when probed by the Hamiltonian \cite{Papadodimas:2017qit}. The Hamiltonian has support on both $\mathcal{A}_R$ and $\mathcal{A'}_R$ and therefore can detect the excitations on the commutant $\mathcal{A'}_R$. Writing the state as $\ket{\psi^e_1} = W(t) \ket{\text{TFD}}$, it can be shown that
\begin{equation}
\frac{d}{dt}\braket{\psi^e_1|\, O(t)\,  H \, |\psi^e_1} =\frac{d}{dt} \braket{\text{TFD}|\, W^{\dagger} \,  O(t) \, [H,W] \, |\text{TFD}} + \text{O}\left(\frac{1}{S}\right). \label{ooe}
\end{equation}

Equation \eqref{ooe} shows that the state $\ket{\psi^e_1}$ is out of equilibrium. The bulk interpretation is now clear as the operators in the right exterior of the black hole cannot detect the excitations $\ket{\psi^e_1}$. These excitations emerge from the past singularity and are short-lived. At around $t \sim t_1$ they arrive at the left part of the diagram. At a later time, they fall into the future singularity. These non equilibrium states are out of equilibrium at around $t \sim t_1$, but remain in equilibrium for $t \ll t_1$ and come back to equilibrium for $t \gg t_1$, and are therefore transient.

It is now easy to generalize from a single excitation to many excitations as given in \eqref{manyexcitations}, where as before, we include the factors $e^{-iH_R T}$'s inside the unitaries $V$'s.
\beq
\ket{\psi^e} = C \, e^{-\frac{\beta H_R}{2}  }\,  V^{\, m}_R (\,t_m) \, V^{\,m-1}_R (\,t_{m-1}) \, \dots \, V^{\,2}_R(t_2) \, V^{\,1}_R (t_1)\, e^{\frac{\beta H_R}{2}  }  \ket{\text{TFD}} \label{manyexcitations2}
\eeq

 This state in \eqref{manyexcitations2} will be seen in equilibrium at $t \gg t_1, \, t_2, \, \dots \, , t_m$ and  $t \ll t_1, \, t_2, \, \dots \, , t_m$. However when probed by the Hamiltonian at intermediate times say at $t \sim t_1, \, t\sim t_2$ or at $t \sim  t_m$ the state will appear out of equilibrium. The bulk picture describing out-of-equilibrium behaviour of the excitations at these intermediate times is understood as them coming out of the past left horizon and travelling in the left exterior before falling into the future horizon (See Figure \ref{eternalparadox}).

 The nature of the excitations reveals the physical picture of the paradox as well. As we have argued earlier, all excitations possess an energy $E_0$, where $m \, E_0 \ll M_{BH}$. This small energy means that the excitations cannot protrude very much outside the interior on the left-hand side, and all excitations protrude a similar distance after coming out of the past horizon before travelling and falling inside the future horizon.

 Now consider early excitations governed by small $t$'s, e.g. $(t_1, \, t_2  \, \dots t_i)$, where $i \ll m$, which come out from the past horizon and fall into the future horizon. These excitations intersect the Cauchy slices with boundaries at earlier Kruskal times and keep intersecting future Cauchy slices at later Kruskal times as well. In contrast, the excitations which come outside the past horizon and fall inside the future horizon at late $t$'s will not intersect the early Kruskal time Cauchy slices. However, these excitations will intersect the late Kruskal time Cauchy slices in the interior (See Figure \ref{eternalparadox}). These above features give rise to the physical picture of the paradox. On the late time slices, there will be more and more excitations where the number of excitations is tuned such that they constitute a dilute gas of a fixed density $\rho$. Therefore we have slices which have an increasingly large value of entropy at late times which becomes more substantial than the Bekenstein Hawking entropy.
 
 \begin{figure}

\centering
\includegraphics[width=.42\textwidth]{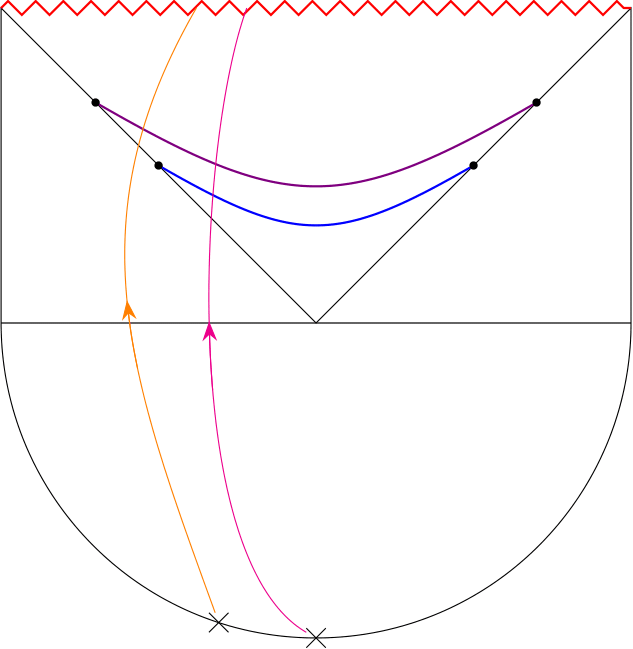}\hfill

\caption{The initial state of the black hole is created by glueing the Euclidean AdS to the bottom half of the Lorentzian Penrose diagram. Boundary deformations of the Euclidean AdS create our excitations. Just after the initial time, all the bags of gold excitations are in the left exterior.}
\label{hh}

\end{figure}

In the bulk Lorentzian description it naively seems that the excitations emerge out of the past singularity. This apparent problem is rectified by writing down an initial bulk state for the problem \cite{Hartle:1976tp, Hartle:1983ai}. The way we construct the initial state or the Hartle Hawking wavefunction of the eternal Lorentzian geometry is by glueing it to a Euclidean AdS part and then performing the path integral over the Euclidean part. We can thus obtain the Hartle Hawking state. At $t = t_0$ all excitations are in the exterior and propagate afterwards on the left side of the Penrose diagram. We write our initial state at this $t = t_0$ when all excitations are outside the horizon.  Here each excitation should be treated as a small deformation of the initial wavefunction and can be generated by inserting operators at the Euclidean AdS boundary as shown in Figure \ref{hh}. This gives us the CFT state \eqref{manyexcitations2}. The initial state in the bulk is a path integral performed over this configuration of an eternal geometry plus small boundary deformations, which is given in \eqref{manyexcitations2}. This path integral qualitatively resolves the problem of constructing a valid initial bulk state in order to pose the paradox.

 \section{The paradox for single sided black holes}
 \label{singleside}
 
 \begin{figure}

\centering
\includegraphics[width=.8\textwidth]{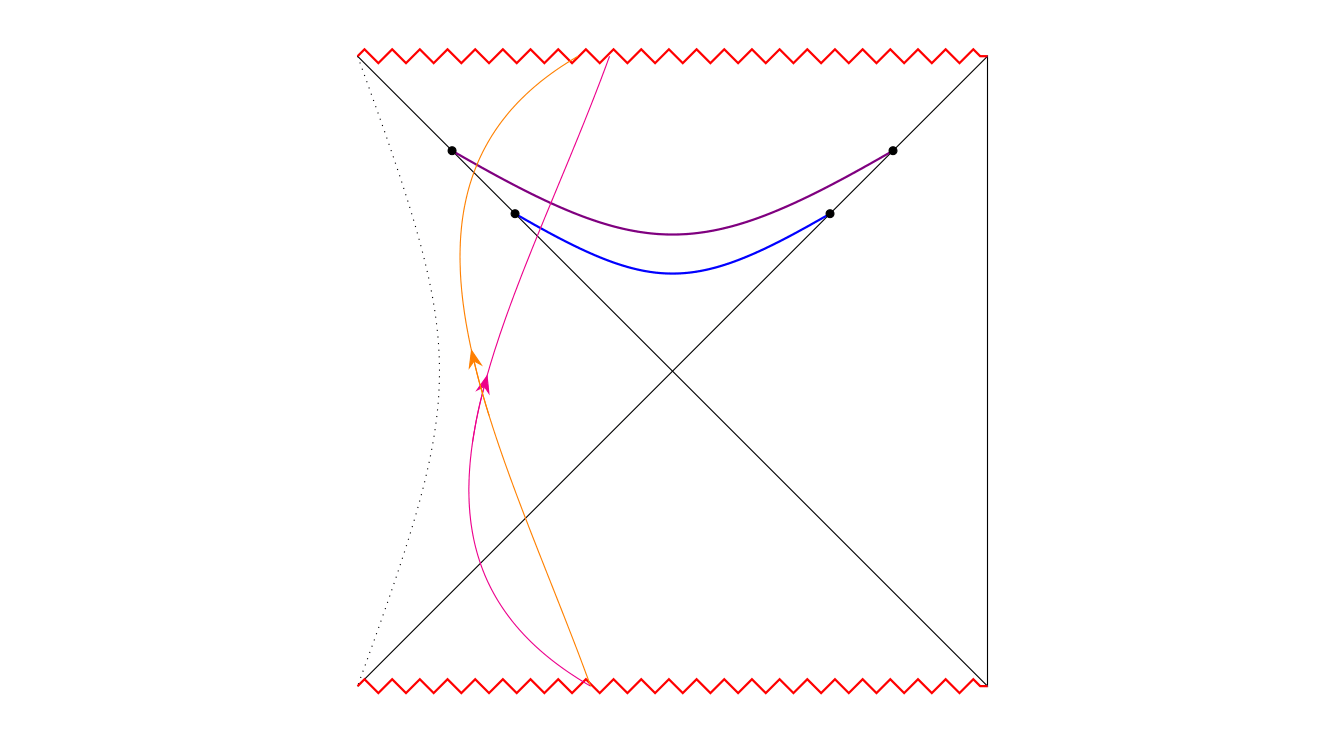}\hfill

\caption{We demonstrate the paradox for pure state black holes which are single-sided. The dotted line on the left denotes the UV cutoff for the theory living on the right boundary, which prohibits us from reaching arbitrary close to the left boundary. Since the interior region is similar for both the single-sided and eternal black holes, the physical picture of the paradox and its resolution is similar.}
\label{pureparadox}

\end{figure} 
 
 Till now, we have discussed at length the paradox for the eternal black hole. We can also pose a similar paradox for pure state black holes. These black holes are described on the boundary by a pure state on a single-sided CFT. Using a single side CFT on the right, the bulk description of the black hole can be reconstructed using HKLL reconstruction \cite{Hamilton:2005ju} in the exterior right \footnote{We thank Debajyoti Sarkar for pointing out related work regarding interior reconstruction \cite{Hamilton:2006fh, Roy:2015pga}.}. The top and bottom regions of the fully extended Kruskal diagram of AdS Schwarzchild black hole can be reconstructed using state-dependent operators as given in Appendix \ref{StateDepReview}. A part of the left bulk region can also be reconstructed; however, we cannot go too far on the left side as one needs operators with higher and higher energies to approach closer and closer to the left boundary. In other words, a UV cutoff on the right boundary CFT prevents us from going arbitrarily close to the left boundary in the bulk.

 Single sided black holes are represented by typical states \cite{lloyd2013pure, PhysRevLett.54.1350, PhysRevLett.80.1373} on the boundary CFT.  We define these states by considering a quantum statistical system at a temperature $T$ and average energy $E$. The relevant example in our case is a CFT at a temperature $T$. Let us consider a small interval $\Delta E$ centred about $E$ in the CFT energy spectrum with $\Delta E \ll E$. We will be looking at $n$ energy eigenstates $\ket{E_i}$ in the interval $\Delta E$, each with energy $E_i$. The entropy is therefore given by $\Delta S = \frac{\Delta E}{T}$, while the number of states $n$ in the interval is related to the entropy as $n = e^{\Delta S}$. We now define a state by randomly superposing the energy eigenstates:
 \beq
 \ket{\psi} = \sum_{i=1}^{n} c_i \ket{E_i}
 \eeq
 such that $\sum_i \abs{c_i}^2 = 1$. Here $c_i$'s are chosen at random. These $ \ket{\psi_s}$ states obey a surprising property: For a quantum statistical system, "almost" all the states $\ket{\psi_s}$ mimic thermal behaviour, and we define such states which look thermal as typical states $\ket{\text{TYP}}$. We pause here to precisely quantify the notion of "almost" and understand how exactly do typical states mimic thermal behaviour.
 
 \subsection*{Statistical properties of typical states}

We will revisit some properties of typical states within a more general formalism. We consider a $n$ dimensional Hilbert space such that $n = \exp S$. Let us work with an orthonormal basis of states $\ket{\psi_i}$ where $i \in [1,n]$. We will now write down the most general pure state living on this Hilbert space:
\beq
\ksi = \sum_{i=1}^{n} c_i \ket{\psi_i} \quad \text{where} \quad \sum_{i=1}^n \abs{c_i}^2 =1. \label{spheretypical}
\eeq

Equation \eqref{spheretypical} describes a sphere $S^{2n-1}$ where all these vectors live, which we previously encountered in \S \ref{howmanyexcitations}. We define the Haar measure on the pure states which guarantees that each pure state is equally likely.

\beq
d\mu = a \, dc_1 \,dc_1^* \,  dc_2 \,dc_2^* \, \dots \,  dc_n \,dc_n^* \, \delta \lc1 - \sum_{i=1}^n \abs{c_i}^2\rc 
\eeq

Here $a$ is fixed using the following condition:

\beq
\int d\mu = 1.
\eeq

 The measure $d\mu$ is invariant under independent rotations of phases $c_i \to e^{i\alpha_i} c_i$. Now consider a linear operator $O$ acting on this Hilbert space $\mathcal{H}$. We want to study the properties of the expectation value $\braket{\psi|O|\psi}$ and also how it depends on $ c_i$'s. We argue that for most choices of $c_i$, the expectation value is independent of the typical state $\psi$ provided that $n$ is very large. Firstly the average $\overline{\braket{\psi|O|\psi}}$ over all states is given by

\beq
\overline{\braket{\psi|O|\psi}} = \int d\mu \, \braket{\psi|O|\psi} = \sum_{i,j} O_{ij} \int d\mu \, c^*_i \, c_j.
\eeq

This integral is non-zero only if $i \neq j$ due to invariance of $d \mu$ under independent rotations of phases $c_i \to e^{i\alpha_i} c_i$. Therefore we write
\beq
\int d \mu \, c_i^* c_j = A_i \, \delta_{ij}, \quad \text{where} \quad A_i = \int d \mu \, \abs{c_i}^2 \label{measure}
\eeq

Since all $c_i$'s enter the measure in an equivalent way and $d \mu$ is independent under permutations of $c_i$'s, the index $i$ on $A_i$ is redundant. Therefore $A_i \equiv A = \int d \mu \, \abs{c_i}^2$. In order to evaluate $A$ we now sum over all $c_i$'s in equation \eqref{measure}. 

\beq
\overline{A} = \int d\mu \, \mathbf{1} = \frac{1}{n}.
\eeq

We use the value of $\overline{A}$ to imply that:
\beq
\int d\mu \, c^*_i \, c_j = \frac{1}{n} \delta_{ij} \implies \overline{\braket{\psi|O|\psi}} = \sum_{i,j} O_{ij} \int d\mu \, c^*_i \, c_j = \frac{1}{n} O_{ij} = \text{Tr} \ls \rho_{\Omega} \,O\rs, 
\eeq
where $\rho_{\Omega} = \frac{1}{n}$ is the microcanonical density matrix. We thus conclude that the the average of the expectation value of operators over all typical states is that of the maximally mixed state. We now want to understand how close is the expectation value of an operator is to the maximally mixed state. In order to do this we need to look at the variance which can be similarly calculated in the following equation:
\beq
\overline{\ls \braket{\psi|O|\psi} - \text{Tr} \,( \rho_{\Omega} \,O) \rs^2} = \frac{1}{e^S +1} \ls \text{Tr}\, ( \rho_{\Omega} \,O^2) - \text{Tr}\, ( \rho_{\Omega} \,O)^2 \rs
\eeq

We see that the variance is exponentially small in entropy. Therefore we conclude that most pure states must look exponentially close to the mixed state or else we will obtain a larger number in the variance. Therefore almost all states mimic thermal behaviour which justifies our claim that almost all states $\ksi$ are typical $\ket{\text{TYP}}$. As a result, we write for almost all such states:
\beq
\braket{\text{TYP}|\,O\,|\text{TYP}} = \text{Tr} \ls \rho_{\Omega} \, O \rs + \text{O} \, \lc \frac{1}{e^S}\rc.
\eeq

An important assumption that goes into calculating the variance is that the degree of the operator $k$ is small compared to the dimension of the Hilbert space $k \ll n$. A violation of this property leads to a more substantial variance. Therefore we demand that the number of operator insertions is much smaller compared to the dimension of Hilbert space. The degree $k$ being small provides a statistical basis for imposing this condition on operators in the small algebra and is the boundary counterpart of demanding that the backreaction due to operator insertions is small.

 Now consider the case where we are looking at energy eigenkets spread over $\Delta E$ such that $\Delta E \ll E$. In this limit the canonical density matrix approaches the microcanonical density matrix. Each typical state therefore satisfies the following property:
 \beq
 \braket{\text{TYP}|\, O_1 \, O_2 \, \dots \, O_k \, |\text{TYP}} = \frac{1}{Z(\beta)} \text{Tr} \ls e^{-\beta H} \, O_1 \, O_2 \, \dots \, O_k \rs + \text{O} \lc \frac{1}{S}\rc \label{typ}
 \eeq
 where we have replaced the microcanonical ensemble with the canonical ensemble. Here again the number of operators $k$ is much smaller than the entropy of the state, $k \ll n$. \eqref{typ} essentially says that $k$-point correlators on the typical state are indistinguishable from thermal $k$-point correlators up to O$\lc\frac{1}{S}\rc$ corrections. This kinematical statement about the correlators is quite surprising; any typical state exhibits such behaviour.

 We will clarify a physical question here. How does the typical state know about the inverse temperature $\beta$? This information is contained in the number of energy eigenstates comprising the typical state, and the energy interval where the states live. Hence the system knows about the temperature.
 
 \subsection{The single sided paradox and its resolution}
 
We now state the paradox for the single-sided black holes. The construction of maximal volume surfaces in \S \ref{maxvolume} is the same for this case because the single-sided black hole possesses the same interior region as the eternal black hole does. The excitations in the black hole interior are also similar with the difference being their action on the typical state rather than on the thermofield double state. A single excitation is given by:

\beq
\ket{\psi^e_1} = C_1 \, e^{-\frac{\beta H_R}{2}  }\,  U_1(t_1) \, e^{\frac{\beta H_R}{2}  }\ket{\text{TYP}} \label{ex2}
\eeq

As before, we can place similar excitations far apart from each other on the nice slice by adjusting the unitary $U$ to create a dilute gas of density $\rho$. Calculating the entropy of this semiclassical configuration again violates the coarse-grained Bekenstein Hawking entropy at late times in the bulk.  The nature of the excitations is also similar, they emerge out from the bottom interior by coming out of the left past horizon and propagate on the left side for some time, and fall into the left future horizon.

Our resolution to the bulk paradox for the single-sided black holes is the same resolution which we have proposed for the eternal case. We have hugely overcounted the excitations in this case as well due to small inner products between coherent bulk states describing the excitations. The resolution for this case is unchanged because the interior possessed by single-sided and eternal black holes is the same.

As before we see that the fine-grained entropy remains unchanged. This consistency arises as the typical state is a pure state and the entanglement entropy of this system is zero. Similar to what was derived in \ref{CFT}, insertion of multiple bulk interior excitations on the typical state leaves the density matrix unchanged. As a result, we again conclude that there is no paradox in the CFT. Even though the fine-grained entropy of the system is zero, the coarse-grained entropy is $S_{BH}$. In the following \S \ref{typicalovercounting} we justify our claim that the enormous number of semiclassical bulk excitations arise due to an overcounting of the bulk Hilbert space.

\subsection{Why interior bulk states are non-orthogonal in the CFT Hilbert space?}
\label{typicalovercounting}

 We consider typical states in the CFT which are dual to the single sided black hole in the bulk and are centered about an average energy $E$ with range $2 \Delta E$:
\beq
\ket{\text{TYP}} = \sum_{i= E -\Delta E}^{E + \Delta E} c_i  \ket{\psi_i}, \label{kettyp}
\eeq
where $\ket{\psi_i}$ are normalized states and $\sum_i |c_i|^2 = 1$. We will denote $\text{O}_{\omega}$ as operators in the boundary CFT with energy $\omega$. \eqref{kettyp} is constructed by acting with a string of $\text{O}$'s on the ground state such that the string's total energy is $E_i$, which then leads to the state $\ket{\psi_i}$. We are looking at states of the form:

\begin{equation}
\ket{\psi} = \text{K} \, e^{-\frac{\beta H}{2}} \, \text{U}(\text{O}_{\omega}) \, e^{\frac{\beta H}{2}} \ket{\text{TYP}}
\end{equation}
where K is the constant of normalization, $\text{U}(\text{O}_{\omega})$ is an unitary operator creating bulk excitation generated by products of $\text{O}_{\omega}$. These operator insertions do not change the energy of the typical state much, i.e. $\omega \ll \text{O}(\mathcal{N})$. Another requirement is that the number of single oscillator operator insertions in $\text{U}(\text{O}_{\omega})$ is lesser than O$(\mathcal{N})$. These conditions define the small algebra of observables $\mathcal{A}$ which act on the ground state to give the small Hilbert space. For the CFT this means that the operator insertions is very small as compared to the energy of the state, and the insertions don't have very high energy themselves. We now want to evaluate the inner product of the two such states in the small Hilbert space, where as previously, $\text{V}_j(\text{O}_{\omega_j}, T_j)$ includes the $e^{iHt}$ insertions, i.e. $\text{V}_j(\text{O}_{\omega_j}, T_j) = e^{iHt} \, \text{U}_j(\text{O}_{\omega_j}) \, e^{-iHt}$.
\beq
\ket{\psi_i} = \text{K}_i \, e^{-\frac{\beta H}{2}} \, \text{V}_i(\text{O}_{\omega_i}, T_i) \, e^{\frac{\beta H}{2}} \ket{\text{TYP}} \quad \& \quad \ket{\psi_j} = \text{K}_j \, e^{-\frac{\beta H}{2}} \, \text{V}_j(\text{O}_{\omega_j},T_j) \, e^{\frac{\beta H}{2}} \ket{\text{TYP}}
\eeq
These states defined above live in the small Hilbert space and the indices $i,j$ go over the small Hilbert space. The inner product between these states is given by 
\beq
\begin{split}
\braket{\psi_j|\psi_i} &= \text{K}_i\, \text{K}_j\braket{\text{TYP}|\, e^{\frac{\beta H}{2}}\, \text{V}^{\dagger}_j(\text{O}_j, T_j) \, e^{-\beta H}\, \text{V}_i(\text{O}_i, T_i) \, e^{\frac{\beta H}{2}}|\text{TYP}}\\
&= \text{K}_i \, \text{K}_j \sum_{k,l,o,p,q,r} c^*_k \,  c_n \braket{E_k|\, e^{\frac{\beta H}{2}} \, |E_o} \braket{E_o|\, \text{V}^{\dagger}_j \, |E_p}\braket{E_p|\, e^{-\beta H}\, |E_q}\braket{E_q|\, \text{V}_i \, |E_r}\braket{E_r|\, e^{\frac{\beta H}{2}}\, |E_l} \\
& = \text{K}_i \, \text{K}_j \sum_{k,l,p} c^*_k \, c_n\, e^{\frac{\beta E_k}{2}} e^{\frac{\beta E_l}{2}} e^{-\beta E_p} \braket{E_k|\, \text{V}^{\dagger}_j\, |E_p} \braket{E_p|\, \text{V}_i\, |E_l} \\ 
\end{split} 
\eeq

We see here that in general, these states are not orthogonal. This non-orthogonality arises since we are working with restricted energy operators on the typical states. Because these states are normalized and since $ e^{\frac{\beta E_k}{2}} e^{\frac{\beta E_l}{2}} e^{-\beta E_p} \approx 1$ as the states lie in the small Hilbert space; we can write the inner products as
\beq
\braket{\psi_j|\psi_i} = \sum_{k,l} (c^j_k)^* \, c^i_l, \label{randomwalk}
\eeq
such that $\sum_l |c^i_l|^2 =1$. Equation \eqref{randomwalk} gives rise to a small but finite  O$(e^{-\frac{S}{2}})$ number, where the dimension of the Hilbert space is $e^S$ \footnote{The derivation of O$(e^{-\frac{S}{2}})$ is straightforward, it is the same as calculating the expected displacement in a random walk problem after $n$ steps.}. We thus see that the inner product between the vectors is a small number if $\ket{\psi_j}$ live in a huge dimensional Hilbert space. These small inner products naturally give rise to overcounting in CFTs. 

\section{Spectral properties of bags of gold spacetimes: Contradictions and Resolution}

\label{spectralsff}

Till now we have discussed the paradox of the coarse-grained entropy of bags of gold spacetimes. Let us now understand the spectral features of these spacetimes in the context of effective field theory. Firstly we will work with the semiclassical Hilbert space of the bags of gold spacetime spanned by the excitations placed far apart from each other. We will argue that such an effective field theoretic description of the Hilbert space potentially contradicts with black holes' spectral observables' predicted behaviour.

Consider the phase space of a classical system exhibiting chaos. It was conjectured in \cite{Bohigas} that the quantum counterpart of such a system should have an energy level spacing distribution which matches one of the three standard random matrix ensembles - Gaussian orthogonal (GOE), unitary (GUE) or symplectic (GSE), depending on the inherent symmetries of the system. Since black holes display scrambling properties, we expect that their level spacing distribution matches the one given by the Gaussian unitary distribution. Therefore a convenient way to model black holes is by using random matrices constructed using Gaussian unitary ensemble.  For GUE, the expression for the level spacing distribution is \cite{mehta2004random}:
\begin{equation}
 P(s) = \frac{32 \, s^2}{\pi^2} \exp \left[ -\frac{4s^2}{\pi} \right], \label{randommatrix4}
\end{equation}
 where $s$ is the distance between two consecutive eigenvalues, which we expect to be the energy level spacing distribution of black holes as well. The conjecture \cite{Bohigas} proposes that the quantum counterpart of a classical system exhibiting chaos possesses either above level spacing distribution or that of its two cousins, the GOE or GSE. This is in contrast to the level spacing distribution obeyed by non-chaotic systems. As a drastically different example, for integrable systems, the Berry-Tabor conjecture states that the level spacing distribution should be Poissonian $P(s) = \exp{-s}$ \cite{10.2307/79349}.

 Using the formalism of random matrix theory, we will argue that the violations of spectral observables can be classified into two types. The nature of the first violation is characterized by qualitative deviation from the expected GUE energy level spacing distribution. To overcome this violation, we will demand that the only bags of gold configurations which are allowed are strictly consistent with a GUE description. Such an imposition drastically constrains the space of allowed bags of gold configurations. We will observe that even after enforcing this condition, bags of gold configurations can still be captured using the spectral form factor; a spectral observable which quantifies the discrete nature of the system. Thus the effective field theoretic description of the bags of gold's Hilbert space suffers from serious contradictions as compared to observed characteristics of black holes. Towards the end of this section, we demonstrate how our overcounting hypothesis resolves these contradictions in the spectral form factor. 

\subsection{Spectral observables in random matrix theory and discrete systems}

\label{sff0}

\subsubsection{Random matrix theory observables}

In this section, we briefly review the spectral observables of random matrices belonging to the Gaussian unitary ensemble. The Gaussian unitary ensemble of Hermitian matrices $H$ of dimension $N \times N$ is defined as follows
\begin{equation}
 Z_{G}= \int [dH] \, \exp\lc -\frac{H^2}{4v^2}\rc. \label{randommatrix1}
\end{equation}

Here $v^2$ is a real number which is O$(1)$ and does not scale with $N$. A convenient way to solve this integral is by decomposing these matrices in terms of their eigenvalues. From \eqref{randommatrix1} the joint probability distribution of the eigenvalues ${\lambda_i, i \in [1,N]}$ belonging to Gaussian unitary ensemble is given by
\begin{equation}
 p (\lambda_1, \dots \lambda_N) = \text{C} \, \exp\left[ 2 \sum_{j<k} \log \abs{\lambda_j - \lambda_i} - \sum_j \frac{\lambda_j^2}{2v^2} \right]. \label{randommatrix2}
\end{equation}

  The first term in the exponential of \eqref{randommatrix2} arises from the Van der Monde determinant, which comes from the Jacobian of the transformation in the measure, while the second term arises due to the Gaussian potential from \eqref{randommatrix1}. The average density of eigenvalues $\bar{\rho}(\lambda)$, where $\bar{\rho}(\lambda) = \int d\lambda_2 \dots d\lambda_N \, p (\lambda, \lambda_2 \dots \lambda_N)$;  is given by
\begin{equation}
 \bar{\rho} (\lambda) = \frac{2 \sqrt{R^2 - \lambda^2}}{\pi R^2}; \quad R^2 = 8v^2 N, \quad -R \leq \lambda \leq R. \label{randommatrix3}
\end{equation}

 Given this setup, we focus on the fluctuations of the eigenvalues, which are independent of the potential $V(\lambda)$ in the large-$N$ limit. Regarding fluctuations, the vital quantity of interest related to quantum chaos is the level spacing distribution given by $P(s)$ as given in \eqref{randommatrix4}.

\subsubsection{Measure of discreteness: Spectral form factor}

As we discussed, apart from the chaotic signatures, since the systems we are studying are black holes which have discrete spectra, it is useful to look at physical observables which can capture discreteness. In this regard, it is useful to understand the typical size of the fluctuations at late times, which in turn characterizes the discreteness of the energy spectrum. In order to define such a quantity, let us first generalize the partition function of a system to include Lorentzian time along with the temperature:
\begin{equation}
 Z(\beta + i T) = \text{Tr} \left[ e^{-\beta H -iHt} \right]. \label{randommatrix5}
\end{equation}

At late times, this generalized partition function oscillates, and the time average of this quantity is zero.  Using this partition function, we will now define the spectral form factor which captures the magnitude of such oscillations:
\begin{equation}
 S(\beta, T) = \frac{Z(\beta + i T) \, Z(\beta - i T)}{Z(\beta)^2} =  \frac{1}{Z(\beta)^2} \sum_{i,j =1}^{N} e^{- \beta (E_i + E_j)}\, e^{i(E_i - E_j)t} \label{randommatrix6}
\end{equation}

 Since the systems in consideration are chaotic, we now demand that the Hamiltonian in consideration is described by a random matrix obeying GUE statistics. Therefore we write the expression for the generalized partition function in Gaussian unitary ensemble:
\begin{equation}
\langle Z(\beta + iT)\rangle_G = \frac{1}{Z_{G}} \int [dH] \, e^{ -\frac{H^2}{4v^2}}\, \text{Tr}\left[ e^{-\beta H - i H T}\right], \label{randommatrix20}
\end{equation}
where $Z_G$ is given by \eqref{randommatrix1}. Equation \eqref{randommatrix20} can now be used to calculate the spectral form factor in \eqref{randommatrix6}. It was shown in \cite{Cotler:2016fpe} that the curve describing the logarithm of spectral form factor versus the logarithm of $T$ obeys the following features:
 
 \begin{enumerate}
  \item The curve starts from 1 and starts decaying with a constant slope at early times. This behaviour can be understood by plugging in the level density in \eqref{randommatrix3} into \eqref{randommatrix20}, and then using it to evaluate the spectral form factor in \eqref{randommatrix6}. The late time decay of the spectral form factor at high temperature is captured by $S(\beta \approx 0, T) \sim T^{-3}$. 
  
  \item
  The decaying behaviour continues until the dip time, after which the curve rises with a constant slope. The physical reason behind this is as follows: $S(T, \beta)$ is roughly a sum of connected and disconnected parts. The disconnected part contributes to the decay which dominates until the "dip time". Equating the late time decay of the spectral form factor and the ramp growth gives the value for the dip time, which is $t_d \sim e^{N/2}$. After dip time the connected part dominates giving rise to the increasing ramp, which at high temperature is given by  $S(\beta \approx 0, T) \sim \frac{T}{2\pi \exp{2N}}$. 
  
  \item
  At a certain time called the plateau time, the ramp stops increasing and gives rise to a constant plateau. Physically the plateau appears because oscillations in the generalized partition function are random and out of phase at very late times, contributing to a small but non-zero number. After the plateau time $t_p> 2e^N$, the constant plateau of the spectral form factor is given by $S(\beta \approx 0, T) \sim \frac{1}{\pi e^N}$.
  
 \end{enumerate}

 This behaviour of the spectral form factor captures the discrete features of black holes, which can be seen from the red curve in Fig. \ref{sfffig2} for $\beta =1$. We will now see that treating the bulk effective degrees of freedom as independent degrees of freedom violates the delicate structure expected from the above description.
 
 \subsection{Spectral properties of bags of gold excitations}
 
  As before, we construct several unitary excitations behind the horizon creating a bags of gold configuration. These excitations are of the form given in \eqref{ex1}, which we restate in the frequency basis:
 \begin{equation}
  \ket{\psi_i} = K_i \,  e^{-\frac{\beta H}{2}} \, \text{V}_i(\text{O}_{\omega_i}, T_i) \, e^{\frac{\beta H}{2}} \, \ket{\psi_{BH}}, \quad i \in (1,m) \label{sff1}
 \end{equation}
 Here as previously, $\text{V}_j(\text{O}_{\omega_j}, T_j)$ includes the $e^{iHt}$ insertions, i.e. $\text{V}_j(\text{O}_{\omega_j}, T_j) = e^{iHt} \, \text{U}_j(\text{O}_{\omega_j}) \, e^{-iHt}$. Here $O_i \in \mathcal{A}$, where $\mathcal{A}$ is the algebra of simple operators. As argued before, these operator insertions have small energies $\omega_i \ll \mathcal{N}=N^2$. Consequently the energies of these excitations belong to a small interval $(E - \Delta E, E + \Delta E)$, where $E \sim O(\mathcal{N})$, and $\Delta E \sim O(1)$.

 In the semiclassical description since states of the form \eqref{sff1} are spread wide apart spatially, we naively think that such distinct configurations have zero inner product. Let us represent the Hilbert space of the effective field theory of the bags of gold spacetime by $\mathcal{H}_{BOG} := \{\ket{\psi_i}\}, \, i \in (1,m)$, which is $m$-dimensional. Following this semiclassical logic, we saw previously that the $m$-dimensional space is very large as compared to the $n$-dimensional black hole's Hilbert space. Since the excitations are placed far apart, this naive reasoning leads us to conclude that the vectors denoting the bags of gold excitations in $\mathcal{H}_{BOG}$ are orthonormal:
 \begin{equation}
  \langle \psi_j\,|\,\psi_i \rangle = 0 , \quad \langle \psi_j|\, e^{iH t} \, |\psi_i \rangle. \label{sff2}
 \end{equation}
 
  \subsubsection{Violations of Type 1}
 
We will now see how this naive EFT description violates the spectral properties expected from \S \ref{sff0}. It is straightforward to construct bags of gold Hilbert spaces spanned by vectors $\mathcal{H}_{BOG} := \{\ket{\psi_i}\}$ such that the difference in the energy levels of these vectors do not obey the expected level spacing distribution given by GUE, which is given in \eqref{randommatrix4}.

\begin{figure}

\centering
\includegraphics[width=.45\textwidth]{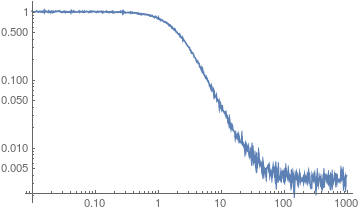}\hfill
\includegraphics[width=.45\textwidth]{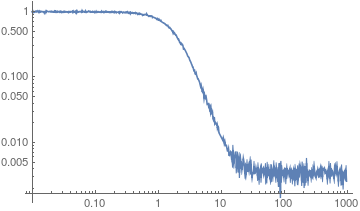}\hfill

\includegraphics[width=.45\textwidth]{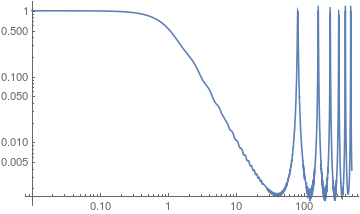} \hfill
\includegraphics[width=.45\textwidth]{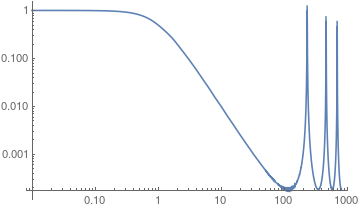} \hfill

\caption{The spectral form factors for different distributions obeying violations of type 1 (See Table \ref{table2}). The figure on the top left consists of random energy levels taken from a uniform probability distribution. The figure on the top right has energy levels picked from a near-uniform probability distribution. Both plots are for 1000 energy levels at $\beta = 2$ over 100 iterations. The bottom figures have uniformly spaced energy levels, and as a consequence, are integrable. The bottom left figure is plotted with 50 energy levels, with $\beta =1$ over 50 iterations, while the bottom right figure is plotted with 250 energy levels, with $\beta =1$ over 50 iterations.}
\label{sfffig1}

\end{figure}

\begin{figure}

\centering
\includegraphics[width=.6\textwidth]{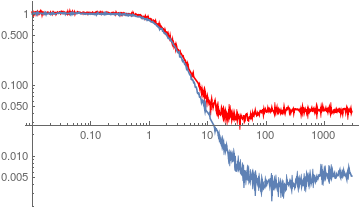}

\caption{The spectral form factors for the "black hole" of 100 states in red and a "bags of gold distribution" with 1000 states in blue plotted using GUE ensemble, over 50 iterations with $\beta =2$. The dip height, dip time, plateau time and plateau height are visible here which are different for both these configurations, which exemplifies violation 2.}
\label{sfffig2}

\end{figure}

A trivial example of such an EFT Hilbert space can be constructed by using vectors of the form \eqref{sff1} such that $V_i\lc O_i (\omega_i),T_i\rc$ has energies $\omega_i$ in integer multiples of a constant $\omega_i = k_i c, \, k_i \in \mathbb{R}$. The above example is an allowed bags of gold configuration because the only physical condition we have enforced is $\sum_i \omega_i \ll O(\mathcal{N})$, with no condition on the individual energies of the excitations. As before, we have denoted the number of black hole states as $n$ and the bags of gold configuration as $m$ with $m \gg n$. Thus the Hilbert space is spanned almost exclusively by the bags of gold states, since $m\gg n$. Therefore in this scenario, the level spacing distribution is that of a bunch of simple harmonic oscillators, which is an integrable system and thus is drastically different from the expected distribution in \eqref{randommatrix4}. In addition, we can see from Fig. \ref{sfffig1} that the spectral form factor does not qualitatively match with the curve expected of black holes. Thus this bags of gold configuration contradicts with spectral features expected from a black hole. 

In general, we can construct various bags of gold spacetimes by spanning the Hilbert space of the EFT using appropriate vectors such that the energy level spacing distribution and the spectral form factor deviates from the spacing distribution and spectral form factor predicted by GUE.  We will call these examples where the energy level spacing distribution and spectral form factor do not qualitatively follow the GUE distribution as \textbf{violations of type 1}.

 \medskip
 
\begin{table}[h!]
\begin{center}
\begin{tabular}{|l|l|l|} 
\hline
 Type & Energy level spacing distribution & Spectral form factor  \\
  & $P(s)$   & $S(\beta, T)$                  \\
  \hline
Violation 1 & No specified distribution     &   No correlation with black hole's curve                            \\
Violation 2 & Follows GUE distribution   & Different $t_d$, $t_p$ and plateau height               \\

\hline
\end{tabular}
\caption{The two types of violations in spectral properties between EFT treatment of bags of gold excitations as independent states and black holes.}
\label{table2}
\end{center}
\end{table}

\begin{table}[h!]
\begin{center}
\begin{tabular}{|l|l|l|l|l|} 
\hline
 Configuration & Dip Time  & Plateau time  & Plateau Height & 2-pt function\\
  & $(t_d) \sim $   & $(t_p) \sim$          &   $\sim$   &   $\langle O(t) O(0)\rangle - G_p \sim$  \\
  \hline
Black hole & $ \sqrt{n}$     &   $n$         &    $n^{-1}$   &     $ t \exp{-2S_{BH}} - \exp{-S_{BH}}$              \\
Bags of gold & $  \sqrt{m}$   & $m$          & $m^{-1}$      &   $ t \exp{-2S_{BOG}} - \exp{-S_{BOG}}$  \\

\hline
\end{tabular}
\caption{Violation 2 - The dip time, plateau time and plateau height for a black hole and a bags of gold configuration in EFT description obeying random matrix statistics in terms of the dimensionality of their Hilbert spaces at $\beta \approx 0$.}
\label{table3}
\end{center}
\end{table}

\subsubsection{Violations of Type 2}

  As seen from the type 1 violations, the effective field-theoretic treatment of bags of gold scenarios can not only lose important features like scrambling etc. but may also result in a completely different description which is integrable. In order to overcome these contradictions, one can demand to consider only those bags of gold spacetimes in which the energy level spacing distribution matches with the GUE level spacing distribution. Such a demand substantially reduces the space of allowed bags of gold spacetimes. Consequently, we have a more refined version of the paradox formulated in the effective field theory Hilbert space which is seemingly consistent with a few basic spectral properties of quantum chaotic systems.

However, we will show that even this restricted space of bags of gold spacetimes which obeys naive GUE level spacing statistics is inconsistent with quantitative features of the spectral form factor involving the height and time of the plateau, dip time and the slope of the ramp, which is due to the fact that $m \gg n$.  We will call these examples where the level spacing distribution follows GUE statistics along with a quantitative deviation from the black hole's spectral form factor as \textbf{violations of type 2}. For convenience, we mention the properties characterizing these two classes of violations in Table \ref{table2}. In order to evaluate the plateau height, we need to look at the long term average of the spectral form factor. The only terms which survive over large times are those with $E_i = E_j$, as the rest of the terms cancel out due to dephasing and thus die off. The long time average of the spectral form factor is thus given by:

\begin{equation}
 \lim_{T \to \infty} \frac{1}{T} \int_0^T dt \, S( \beta, t) = \frac{1}{Z(\beta)^2}  \sum_{i =1}^{N} g^2(E_i) e^{-2\beta E_i} = \frac{1}{Z(\beta)^2}  \sum_{k =1}^{n} e^{-2\beta E_k}, \label{randommatrix7}
\end{equation}
where $g (E_i)$ denotes the degeneracy of states at energy $E_i$. For the EFT description of the bags of gold spacetime, the plateau height is given by:

\begin{equation}
 \lim_{T \to \infty} \frac{1}{T} \int_0^T dt \, S( \beta, t) = \frac{1}{Z(\beta)^2}  \sum_{i =1}^{N'} g^2(E_i) e^{-2\beta E_i} = \frac{1}{Z(\beta)^2}  \sum_{k =1}^{m} e^{-2\beta E_k}, \label{randommatrix8}
\end{equation}

Here $m$ is the number of the bags of gold states, such that $m \gg n$. Since $m \gg n$ there is a quantitative disagreement between the plateau height of the original black hole and the bags of gold spacetime. For the high temperature case with $\beta \approx 0$ as described in the \S \ref{sff0}, we can conclude that the plateau height is $e^{-N} = \frac{1}{n}$ for the original black hole and $e^{-N'} = \frac{1}{m}$ for the bags of gold spacetime. In addition the dip time is $t_d \sim e^{\frac{N}{2}}$ for the black hole and $t_d \sim e^{\frac{N'}{2}}$ for the bags of gold respectively, while the plateau time is $t_p \sim e^N $ for the original black hole, while $t_p \sim e^{N'}$ for the bags of gold respectively. These values are collectively summarized in Table \ref{table3}. Thus even if we choose the bags of gold configurations in such a way that they obey naively obey qualitative spectral properties, there are quantitative differences which are captured using the spectral form factor.

\cite{Cotler:2016fpe} also pointed out the behaviour of the two-point function with the assumption that the system obeys the eigenstate thermalization hypothesis, and has a ramp at late times. They predicted that the two-point function should be of the following form:
\begin{equation}
 G(t) = \langle O(t) O(0)\rangle \sim G_p + \frac{t}{L^2} - \frac{1}{L},
\end{equation}
where $L \sim \exp{S}$ of the system. Thus the two-point function for the effective field theory of bags of gold and the black hole has different behaviour, as mentioned in Table \ref{table3}.

 \subsection{Resolution of spectral puzzles using overcounting}

We now ask whether our earlier proposed resolution to the paradox reconciles these disagreements. We study the spectral properties in the context of pure state black holes for convenience, and we are interested in the order of magnitude of the partition function. Our conclusions can be extrapolated to general black holes as well. As before, we consider typical states defined on the interval $(E - \Delta E, \, E + \Delta E)$ to represent a pure state black hole. The partition function of the dual CFT describing the original black hole over this interval has the order of magnitude:
\begin{equation}
 Z(\beta) = \text{Tr} \left( e^{-\beta H}\right) \sim O(n e^{-\beta E}). \label{sff3}
\end{equation}

Here we have considered $\Delta E \ll E$, which gives us the above order of magnitude of the partition function. We now evaluate the partition function of the bags of gold case where we assume that the $m$ states spanning the EFT Hilbert space are orthogonal. Therefore the order of magnitude of the partition function is given by:
\begin{equation}
 Z_{BOG} (\beta) = \text{Tr} \left( e^{-\beta H}\right) = \sum_{l =1}^{m} \sum_{i,j = 1}^{n} \langle l|i\rangle\langle i|e^{-\beta H} |j\rangle \langle j|l\rangle = O(m e^{-\beta E}). \label{sff4}
\end{equation}

This overcounting in the partition function manifests itself in wrong quantitative values for the entropy of the black hole, spectral form factor and the two-point function at late times. As earlier, we will argue that the bags of gold states in quantum gravity are not independent but have small inner products with each other. Therefore the actual Hilbert space is spanned by $n$ vectors, with $m$ embedded Bags of gold vectors which have tiny but non-zero inner products between each other. Thus we can arrive at the correct conclusion that $Z_{BOG} \sim O(n e^{-\beta E})$ by working with bulk states such that they have small but finite inner products. The above conclusion holds as the correct sum over $l$ in \eqref{sff4} is really up to $n$ instead of up to $m$.  The conclusion that $Z_{BOG} \sim O(n e^{-\beta E})$ is also consistent with the entropy of the black hole as seen before. Similarly we repeat this analysis for $Z(\beta, T)$ as well, and hence we argue that the correct spectral form factor for the bags of gold spacetime should match the black hole's spectral form factor by thinking about the bulk interior states as embedded in the $n$-dimensional Hilbert space with small inner products.

Another way to verify that overcounting resolves discrepancies is by observing that the spectral form factor for the bags of gold configuration quantitatively matches with the black hole's spectral form factor in Table \ref{table3} if overcounting is taken into account. Given that the actual dimensionality of the $m$ dimensional overcounted Hilbert space is $n$, we see that the dip time, the plateau time and the height of the plateau for the bags of gold configurations match with the original black hole's curve's features. Similarly, overcounting resolves the discrepancy between the 2-pt function in the bags of gold spacetime and the black hole as well, in accordance with our earlier argument that Bekeknstein-Hawking entropy gives the correct entropy of bags of gold configurations.

 \section{Study of the paradox using toy matrix models}
 \label{matrixmodel}
 
  In this section we explicitly demonstrate how overcounting allows us to construct an immense number of bulk excitations in the context of toy matrix models. Even though the bulk states arise from matrix models in the large $N$ limit, we can understand aspects of overcounting by performing computations even in small $N$ toy matrix models. Such matrix models have a small dimension of the Hilbert space, and it is possible to list out the state space explicitly.

  By calculating the partition function of a matrix model at temperature $T$ using the canonical ensemble, we can extract out the average energy and entropy of the system. Exponentiating the entropy gives us the dimension of the Hilbert space. We are interested in the regime of small $N$ and temperature such that the dimension of the Hilbert space is less than $1000$.

 We will demonstrate overcounting in two different toy matrix models. The first example is of a $(0+1)$ dimensional two matrix model which has a $U(N)$ global symmetry group. We construct a typical state using the microcanonical ensemble. Afterwards, we will write down the small Hilbert space and demonstrate that we can embed a larger number of vectors compared to the dimension of the small Hilbert space. The second example deals with a CFT consisting of 2 matrices defined on $S_3 \times \mathbb{R}$. Here we will calculate the Hagedorn temperature and construct the typical state above the Hagedorn temperature. Again we will construct the small Hilbert space and demonstrate overcounting. These toy examples show that overcounting with small inner products is natural in the small Hilbert space. 
 
 A similar construction of states follows for bulk states at large $N$.  Apart from computational problems with enumerating the states explicitly, there is no further restriction to doing the same for CFTs with holographic duals in the large $N$ limit. Some results obtained in this section are in agreement with recent related work  \cite{Milekhin:2020zpg}.

\subsection{Toy Model I: A $(0+1)$-d two matrix model}

We will work with the two matrix model given by
\beq
L = \frac{1}{2} \text{Tr} \ls \lc \del_{t} A'_{ij} \rc^2 - \omega_A^2 \lc  A'_{ij} \rc^2 + \lc \del_{t} B'_{ij}  \rc^2 - \omega_B^2 \lc  B'_{ij} \rc^2 + \lambda A'_{ij}  B'_{ij} \rs.
\eeq
Here we have put the interaction term with coupling $\lambda$ so that $A'$ and $B'$ are not independently diagonalized. We also demand that $\lambda \approx 0$, so that this coupling term does not have a significant contribution to the energy and the low energy states are the same as the states in the free field theory to a very good approximation. The $\omega$'s here enforce an IR cutoff, and as a result, we do not have any soft modes in the problem.

 We will also impose that our physical observables are singlets of the global group \footnote{In a sense this replicates features of matrix models in which the matrices transform under a gauge group, where the relevant observables are gauge singlets.}. Therefore we diagonalize $B_{ij}$ by using $U^{-1} B' U = B$, with $B$ being a diagonal matrix comprising of the eigenvalues of $B$. Under the same transformation, $U^{-1} A' U = A$ which is a non-diagonal matrix. Note that this transformation is akin to gauge fixing and a similar transformation in gauged matrix models removes most of the gauge freedom. In the limit $\lambda \approx 0$, the equations of motion of $A$ and $B$ are given by 
\beq
\ls \frac{\del^2}{\del t^2} + \omega_A^2 \rs A_{ij}(\vec{x}, t) = 0, \quad \ls \frac{\del^2}{\del t^2} + \omega_B^2 \rs B_{ii}(\vec{x}, t) = 0.
\eeq

In this case, we will have $N^2 + N$ number of independent oscillators, with $N^2$ coming from $A$ and $N$ coming from $B$. In the large $N$ limit, the $N^2$ oscillators are responsible for the Hagedorn growth of states. We will quantize the system by imposing the commutation relations

\beq
\ls a_{ij}(k), \,  a^{\dagger}_{i'j'}(k')  \rs =  \delta_{ii'} \, \delta_{jj'} \quad \ls b_{ii}(k), \,  b^{\dagger}_{i'i'}(k')  \rs =  \delta_{ii'}  \, \delta_{ii'}.
\eeq

The vacuum of this system is given in the following equation. We generate the state space by the repeated application of these oscillators on the vacuum.
\beq
a_{ij} \ket{0} = 0, \quad b_{ii} \ket{0} = 0.
\eeq

\subsubsection{Typical states, the small algebra and the small Hilbert space}

\begin{figure}

\centering
\includegraphics[width=.49\textwidth]{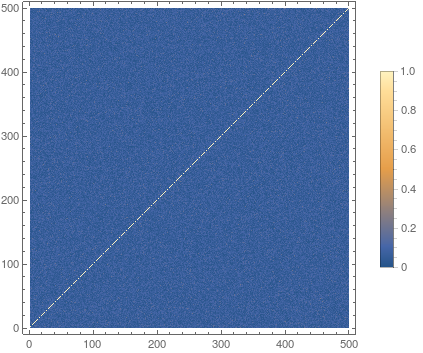}\hfill
\includegraphics[width=.49\textwidth]{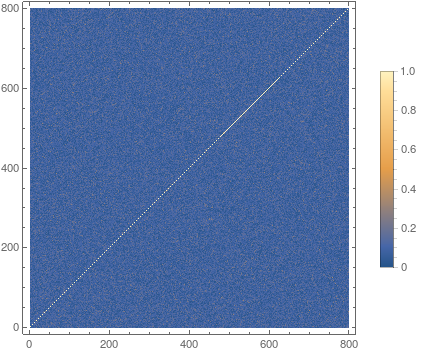}\hfill

\caption{We display the inner products which arise from embedding $m = 500$ approximately equidistant vectors in the $n = 104$-dimensional small Hilbert space of Toy Model I in the left figure and $m=800$ vectors in the same Hilbert space in the right figure. A point on these plots corresponds to the absolute value of the inner product between vectors lying on the $x$-axis and the $y$-axis and hence $x=y$ line has inner product equal to 1.}
\label{static1}

\end{figure}

We work with $N = 4$ for Toy Model I, for which we have $N^2 +N = 20$ creation and annihilation operators. We will set the zero-point energy of the matrix model to zero for our case by subtracting it off from the energy and thus redefining it, and set $\omega_A$ and $\omega_B$ both to 1 while setting $\lambda = 0.01$.

The first thing to construct here is the typical state. To do this we first select energy eigenstates in a range $\Delta E$ about average energy $E$ such that the energies lie in the interval $E \pm \frac{\Delta E}{2}$. The typical state is now created using a random superposition of these energy eigenstates. We take the $E = 16$ with an interval $\frac{\Delta E}{2} = 3$. We now construct a typical state with random $c_i$'s weighing energy eigenstates in the interval $\Delta E$ such that $\sum_i \abs{c_i}^2 = 1$. The inverse temperature $\beta$ of this system is calculated using the first law $\beta = \frac{\Delta S}{\Delta E} = 0.92$.

We will now construct the small algebra and subsequently, create the small Hilbert space. We will demand the following three conditions on the small algebra:

\begin{itemize}
 \item None of the operators in the small algebra annihilates the typical state.
 \item The maximum number of operator insertions on the state is less than 20, i.e. should be lesser than O$(N^2+N)$. We take the maximum number 4.
 \item The maximum energy of the operator insertions is $3$, i.e. much less than average energy $E$ which in our case is 16. The energy of operator insertions should not take us outside $\Delta E$ about $E$ in order to ensure that the backreaction is small.
\end{itemize}

\begin{figure}

\centering
\includegraphics[width=.49\textwidth]{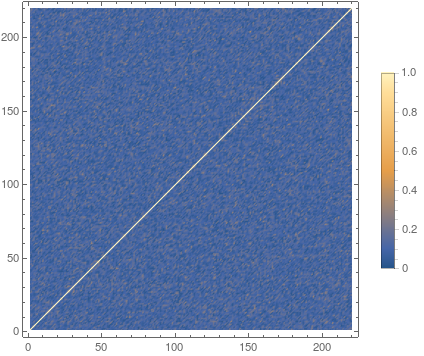}\hfill
\includegraphics[width=.49\textwidth]{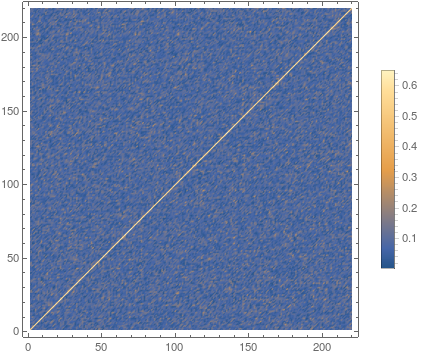}\hfill
\includegraphics[width=.49\textwidth]{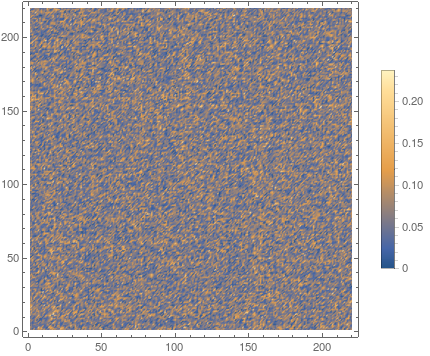}\hfill

\caption{We create $m=220$ excitations in the small Hilbert space of Toy model I. Each excitation is separated from the previous one by time $T=0$ in the figure on the top left, $T=1000$ in the top right and $T=10^5$ in the bottom. We see that increasing time separation gradually washes out the inner products, especially the correlations on the line $x=y$. This behaviour of the inner products indicates that spatially separated excitations on the maximal volume slice have small inner products and hence a "fat tail" in quantum gravity and deviates from the semiclassical zero overlap prediction.}
\label{dynamic}

\end{figure}

Using the above conditions we can identify all 104 possible operators and act them on the typical state to generate the 104-dimensional small Hilbert space.
 
 \beq
 \mathcal{H}_{ \, \ket{\text{TYP}}} := \mathcal{A}  \, \ket{\text{TYP}},
 \eeq
 Although not orthogonal these vectors are all linearly independent. As a cross check, we computed the rank of the matrix constructed with all these vectors, which was found to be 104. 
 
 \subsubsection{Kinematical demonstration of overcounting in toy model I}
 
 We will now use the small Hilbert space to create the interior states as given in equation \eqref{vectors2} where $\text{O}_i(\omega) \in \mathcal{A}$.

\beq
\ket{\psi_i} = \text{K}_i \, e^{-\frac{\beta H}{2}} \, \text{O}_i(\omega) \, e^{\frac{\beta H}{2}}  \, \ket{\text{TYP}} \label{vectors2}
\eeq

 We now construct interior "bulk-like states" by taking combinations of singlet states in \eqref{vectors2}. Each of these states corresponds to the action of an "interior bulk" operator on the typical state. We generate $m=220$ vectors $\ket{v^j}$  spaced apart from each other in the Hilbert space by defining an energy cost between them, which minimizes their inner products. We implement this energy cost numerically by pushing the vectors around in the small Hilbert space (the sphere discussed in \S \ref{howmanyexcitations}) such that they roughly become equidistant. We discuss this technique in detail in Appendix \ref{technique}.  The resulting "interior bulk states" are given below where each of them depends on the choice of coefficients $Z^j_i$
\beq
\ket{v^j} = \sum_i Z^j_i \ket{\psi_i}. \label{smearedbulk}
\eeq
Here the choice of $Z^j_i$ is determined by the energy cost which we can manually select. We plot the vectors' inner products in Figure \ref{static1}, where each point denotes the absolute value of the inner product between a vector on the $x$-axis and a vector on the $y$-axis. The $x=y$ line has inner product 1, which indicates that these vectors are normalized. As a consistency check the $220 \times 220$ matrix generated by these "bulk states" has rank 104. It can be seen from Figure \ref{static1} that there is a finitely non-zero inner product between these bulk vectors. As we increases the dimension of the Hilbert space, these inner products can be made quite small yet finite.

 \subsubsection{Excitations separated far apart in the "interior"}

Till now we have given a kinematical description of the "bulk excitations", i.e. we took the Hilbert space and showed that there exist vectors which have small inner products. In order to model the placement of the "bulk excitations" far apart on the maximum volume slices of the black hole we need to send in each excitation long after the previous one. The static description corresponds to the excitations all sent in at the same time, which means that independent excitations are lying nearby close enough on the Cauchy slice and are not separated far apart. We now plot the dynamical case in Figure \ref{dynamic} where we send subsequent excitations with a time $T$ separated between them. We now model the bags of gold paradox as given in \S \ref{maxvolume} by placing the  excitations far apart from each other on the Cauchy slice which corresponds to large numerical values of $T$.

We note a few interesting observations regarding the dynamical plot. The diagonal line here is the inner product of a vector on the $x$-axis with time evolution acting on the same vector on the $y$-axis. At $T=1000$, we see that the diagonal line fades away a bit and the larger inner products get slowly washed out. At a very late time, $T = 10^{5}$ the diagonal line completely vanishes. This disappearance corresponds to the case when the excitations on the bulk are placed quite far apart on the maximal volume slices. As we can see, the time evolution washes out correlations between the vectors, and the larger inner products cease to exist. Such a washing-out behaviour verifies the "fat tail" of inner products which means that at late times the CFT excitations have a small overlap and is consistent with our derivation in \S \ref{typicalovercounting}. This numerical overlap becomes lesser and lesser if the dimension of the Hilbert space increases because there is much more space in the Hilbert space to accommodate all the vectors.

\subsection{Toy Model II: A $(3+1)$-d CFT on $S_3 \times \mathbb{R}$}

Toy matrix model I illustrates basic overcounting features for a thermal state constructed out of a matrix model. We will now proceed onto another example which is  given by a CFT toy model. Here we first write down the CFT partition function and use it to calculate the Hagedorn temperature which allows us to work in the regime of big AdS black holes. We will construct a typical state at a temperature just above the Hagedorn temperature and demonstrate overcounting of bulk excitations. The metric on $S_3 \times \mathbb{R}$ is given by:
\beq
ds^2 = - dt^2 + a^2 \, d\psi^2 + a^2 \sin^2 \psi \lc d\theta^2 + \sin^2 \theta \, d\phi^2\rc, \label{CFTmetric}
\eeq
where $a$ is the radius of the $S_3$; $\psi$ and $ \theta$ go from $\lc 0, \pi\rc$ and $\phi$ goes from $\lc 0, 2\pi\rc$. On this manifold, we write down a CFT action of two matrix-valued bosonic oscillators $A$ and $B$ transforming under the adjoint representation of $U(N)$ global group in \eqref{toymodel2}. 
 
 \beq
 S_{\text{CFT}}  = \text{Tr} \ls- \frac{1}{2} \int d^4 x \sqrt{-g}\,  \lc  g^{\mu \nu} \partial_{\mu} A'_{ij}\,  \partial_{\nu} A'_{ij}  + g^{\mu \nu} \partial_{\mu} B'_{ij}\,  \partial_{\nu} B'_{ij} + \frac{R}{6} \ls (A'_{ij})^2 + (B'_{ij})^2 \rs  \rc \rs \label{toymodel2}
 \eeq
 
 For the metric given in \eqref{CFTmetric} the Ricci scalar is given by $R = \frac{6}{a^2}$. As in the previous toy model, we will add a small interaction term with a coupling $\lambda \approx 0$. The small coupling ensures that matrices $A'$ and $B'$ cannot be diagonalized independently, and the energy eigenstates are approximately the same as that of free matrix models.
 
 \beq
 S = S_{\text{CFT}} - \text{Tr} \ls \frac{\lambda}{2} \int d^4 x \sqrt{-g}\,  A'_{ij} B'_{ij}  \rs
 \eeq
 
 We will again demand that the physical observables are global group singlets. This time instead of fixing the $U(N)$ matrices using diagonalization, we will perform a precise counting of the number of global group singlets constituting a thermal ensemble. We are interested in the following physical observables: average energy, entropy and the dimensionality of the Hilbert space. We will derive these quantities by evaluating the thermal partition function of the matrix model. We outline this calculation in Appendix \ref{partitionfunction} where we count the number of group singlets using characters of $U(N)$ group and use it to write down the partition function in terms of a Coulomb gas problem with an attractive and a repulsive term. Counting only the group singlets allows us to model the confinement-deconfinement phase transition in the matrix model \cite{Aharony:2003sx, Sundborg:1999ue, Witten:1998zw}. We calculate that the "Hagedorn temperature" of this system is given by $T_H = 0.63$. The thermodynamic observables at a temperature slightly above Hagedorn temperature $T = 0.64$ are listed in Table \ref{table}.

\begin{table}[h!]
\begin{center}
\begin{tabular}{|l|l|l|l|} 
\hline
N & Entropy & Average energy & Dimension of Hilbert space \\
  & $(S_N)$   & $(E_N)$          &  $(D \approx e^{S_N})$        \\
  \hline
2 & 3.43    & 0.54           & 31                         \\
3 & 4.76    & 1.64           & 116                        \\
4 & 5.68    & 3.25           & 293                        \\
5 & 6.59    & 5.63           & 725                      \\  
\hline
\end{tabular}
\caption{Entropy, Average energy and dimension of Hilbert space for small $N$ Toy Matrix model II at $T = 0.64$}
\label{table}
\end{center}
\end{table}
 
 \subsubsection{Typical states, the small algebra and the small Hilbert space}
 
 We will work with the $N=5$ case, which gives us $N^2 + N=30$ independent oscillators. We set the following parameters: the radius of $S_3$ is given by $a =1.55$, $\lambda = 0.01$ and $T = 0.64$. As in the previous model we now construct a typical state with $T = 0.64$, which we accomplish by taking states in an interval $\Delta E$ such that $\frac{1}{T} = \frac{\Delta S}{\Delta E}$. We take energy eigenstates spreaded within $\Delta E = 3$ about $E = 5.63$ and create the typical state by random superposition of these vectors. This gives us a microcanonical description of the matrix model for $N=5$ at $T = 0.64$, the canonical description of which is given in Table \ref{table}.

 We again construct the small Hilbert space by the action of the small algebra on this typical state, where the small algebra satisfies the following conditions:
 
 \begin{itemize}
  \item The number of operator insertions on the state is much lesser than $30$, i.e. should be lesser than O$(N^2+N)$. We choose that the maximum number of operator insertions on the typical state is 1.
  
  \item
  The maximum energy of the operator insertions is $1.5$, i.e. much less than average energy $E$ which in our case is 16. The energy of operator insertions should not take us outside $\Delta E$ about $E$ in order to ensure that the backreaction is small.
  
   \item None of the operators in the small algebra annihilates the typical state.
 \end{itemize}

 \begin{figure}

\centering
\includegraphics[width=.49\textwidth]{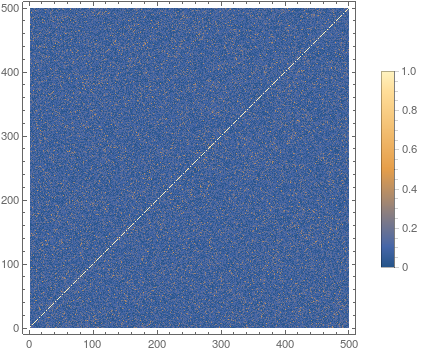}\hfill
\includegraphics[width=.49\textwidth]{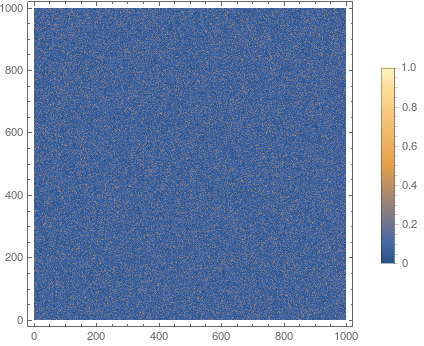}\hfill

\caption{In the left figure, we have $m=500$ excitations embedded in the $n =61$-dimensional small Hilbert space of Toy model II. This embedding corresponds to the static case where the time difference between consequent excitations is $T=0$. The right figure deals with $m =1000$ excitations created from the $n =61$-dimensional Hilbert space. Here we have a time difference of $T = 10^5$ between consecutive excitations. This case corresponds to inner products between excitations placed far apart on the maximal volume slice. The inner products saturate at late times which we can see from the washing out of inner products on the $x=y$ line.}
\label{static2}

\end{figure}
 
 Since the maximum number of operator insertions is 1, we have $61$ states generated by the creation and annihilation operators, and hence the dimension of the small Hilbert space is 61. We will now construct the interior states using operators $\text{O}_i(\omega) \in \mathcal{A}$:
 \beq
\ket{\psi_i} = \text{K}_i \, e^{-\frac{\beta H}{2}} \, \text{O}_i(\omega) \, e^{\frac{\beta H}{2}}  \, \ket{\text{TYP}} \label{vectors1}
\eeq

 \subsubsection{Kinematical demonstration of overcounting in toy model II}
 
The vectors in \eqref{vectors1} constitute the interior bulk excitations in the large $N$ limit, where smeared semiclassical states correspond to combinations of these excitations living in the small Hilbert space. These bulk excitations have the form given in equation \ref{smearedbulk}. Using the energy cost defined in Appendix \ref{technique}, we now construct $m=500$ bulk excitations as in Figure \ref{static2}, which are approximately equidistant from each other.  Each point in Figure corresponds to the inner product's absolute value between a vector on the $x$-axis and the $y$-axis. As expected, the $x=y$ line has an inner product of 1 along it since the states are normalized.

  \subsubsection{Excitations separated far apart in the "interior"}

Till now, we have analyzed overcounting for the static case where the excitations are all situated close to each other on the maximal volume slices in bulk. We now proceed to the dynamical case where we separate the excitations in time, and the corresponding bulk states are spatially separated far apart from each other on the maximal volume nice slices. We expect from the previous toy model that the inner products between these excitations get washed out at huge time separations. We now model $m=1000$ vectors embedded in the 61 dimensional Hilbert space and confirm this in Figure \ref{static2}. Here we see that the inner products saturate at a minuscule value if the time difference between two successive excitations is $T = 10^5$. Thus we obtain the predicted "fat tail" for this CFT as well. 

\medskip

We have thus shown using two toy matrix models that the inner products between excitations spaced far apart on the maximum volume slices deviate away from the semiclassically predicted inner product. This is consistent with our finding in \S \ref{typicalovercounting} that the inner products in CFTs get saturated at a small number. These serve as examples demonstrating our resolution in \S \ref{resolution} that the inner products are essential to resolve the bags of gold paradox.
  
  \section{General properties of systems with overcounted Hilbert spaces}
  \label{genproperties}
  
  We pose the following important question in this section: Since our proposed resolution says that the Hilbert space is overcounted due to small inner products between vectors, what are the physical consequences of such a resolution? In other words, can physical systems in our real-world also have a similar overcounting situation, thereby leading to a much smaller Hilbert space than what we think they have? We see that there exist some significant obstructions to such a situation.
  
  \subsection*{Simulating a quantum system's Hilbert space using a smaller Hilbert space}
 Consider an $m$-dimensional "original" Hilbert space, which can be spanned by $m$ orthonormal vectors. We will now simulate the $m$-dimensional Hilbert space using a smaller $n$-dimensional Hilbert space, such that $m > n$ and see whether it leads to any inconsistency in physical observables. We construct a nearly orthogonal basis of $m$ vectors, which then "spans" the larger space with the following inner products:
\beq
\langle V_1|V_m\rangle=0,\quad \langle V_1|V_i\rangle=\langle V_m|V_i\rangle=\epsilon \quad \forall i=2,\ldots,m-1; \quad \quad \langle V_i|V_j\rangle=0 \quad \forall \, i,j=2,\ldots,m-2
\eeq

These vectors are simulating orthogonal states in the larger Hilbert space. We will now consider the Hamiltonian acting on the $m$-dimensional space given by:
\beq
H=\sum_{i=2}^{m-1}|V_i\rangle\langle V_i|. 
\eeq

This Hamiltonian time evolves the state $\ket{V_1}$ to $e^{-iHt} \, |V_1\rangle.$ Therefore starting from $|V_1\rangle$, time evolution will never lead to $|V_m\rangle$ in the original Hilbert space. We will keep the form of the Hamiltonian same in the smaller Hilbert space in order to not tamper with the energy spectrum. This time evolution takes place within the  space spanned by the vectors:
$$
|V_1\rangle, \quad  |\Psi\rangle=\frac{1}{\sqrt{m-2}}P\sum_{i=2}^{m-2}|V_i\rangle, \quad |V_m\rangle,
$$
where $P$ denotes a projector which projects a vector onto the subspace orthogonal to $|V_1\rangle$ and $|V_m\rangle$. In this basis, the Hamiltonian in the above subspace takes the form:

\beq \label{overcountedmatrix}
\approx\left(\begin{array}{ccc}
m\epsilon^2 & \sqrt{m}\epsilon & m\epsilon^2 \\
\sqrt{m}\epsilon & 1 & \sqrt{m}\epsilon \\
m\epsilon^2 & \sqrt{m}\epsilon & m\epsilon^2 
\end{array}\right).
\eeq

The original Hamiltonian acting on the $m$-dimensional Hilbert space expressed in the orthonormal basis is given by the following matrix,
\beq
\left(\begin{array}{ccc} 0 & 0 & 0 \\ 0 & 1 & 0 \\ 0 & 0 & 0 \end{array}\right).
\eeq

These matrices are not the same and their physical properties are very different for large $m$. The matrix given in equation \eqref{overcountedmatrix} can almost perfectly transfer the state $|V_1\rangle$ to $|V_m\rangle$ in a time $t =\pi/(1+2m\epsilon^2)$ \cite{PhysRevLett.91.207901, PhysRevA.71.032312, doi:10.1142/S0219749910006514}. Therefore we arrive at a contradiction here. If we try to simulate a system without compromising upon the Hamiltonian's form, then they can behave erratically under time evolution. Conventional quantum systems thus cannot be described using a smaller Hilbert space as they can demonstrate forbidden quantum state transfers. Such quantum state transfers are a generic feature of simulated larger Hilbert spaces. Earlier, we argued that the semiclassical Hilbert space of gravity is a simulated Hilbert space with small inner products. It will be interesting to understand precisely what kind of such quantum state transfers occur in semiclassical gravity, and what novel physical features do they display.

\subsection*{Simulating a thermal system using a smaller Hilbert space}

Consider vectors in an $n$ dimensional Hilbert space simulating a larger $m$ dimensional Hilbert space with $m \gg n $. Here we consider that the physical system is thermal. The vectors in the Hilbert space satisfy the following conditions:
 \begin{equation}
\braket{v_i|v_i} =1 \quad \& \quad \abs{\braket{v_i|v_j}} \sim \epsilon, \quad i \neq j. 
\end{equation}

The thermal system under consideration is specified by energy levels spreaded over $\lc E \pm \Delta E\rc$ such that $\Delta E \ll E$. We are interested in the order of magnitude of the partition function, which is given by:

\beq
Z(\beta) = \text{Tr}\lc e^{-\beta H}\rc = \text{O}\lc m  \,e^{-\beta |E|}\rc.
\eeq

We will now see that the simulated thermal system's partition function is has a significant correction.

\beq
Z_{\text{sim}}(\beta) = \text{Tr}\lc e^{-\beta H}\rc \sim \sum_i e^{-\beta E_i} + \sum_{i \neq j} \ls e^{-\beta E} \epsilon^2\rs_{ij}= \text{O}\ls m \, e^{-\beta |E|} \, (1 + m \epsilon^2 )\rs.
\eeq  

We can see that even with tiny corrections to the inner product of the order of $\abs{\epsilon} \sim \frac{1}{\sqrt{m}}$ we will end up with an immense contribution to the partition function. Thermodynamic observables in a system are functions of the partition function and its derivatives. It is safe to say that such a significant contribution to the partition function messes up details of the thermodynamic observables in the system.

As long as $\abs{\epsilon} \ll \frac{1}{\sqrt{m}}$ we don't have a problem with the thermodynamic observables. This is consistent with our observation from \eqref{ned} that such a situation does not lead to the possibility of a big overcounting.

\section{Discussion and conclusions}

 In our work, we have demonstrated a possible resolution to understand the case of several excitations living in the black hole interior. We have proposed that these numerous excitations living on large volume Cauchy slices in the interior are not inconsistent with the Bekenstein Hawking entropy, as they have small inner products and thus are not independent excitations. We advocate that such a situation is not a problematic feature of effective field theory but an essential aspect of quantum gravity. This overcounting naturally arises in the context of boundary theories as shown using toy matrix models. We also showed that spectral observables like the form factor and the level spacing distribution are violated in the semiclassical treatment of the interior excitations, and our proposal resolves this contradiction as well.

 The notion of small inner products is consistent with the breakdown of locality in quantum gravity. We expect locality to hold in effective field theories. In contrast, in quantum gravity, we expect locality to hold approximately. Here locality can break down in various situations, such as the case where we act with too many probes on the spacetime \cite{Ghosh:2017pel}. Small inner products between spatially separated excitations can be understood to be another such situation which demonstrates the breakdown of locality in quantum gravity.

 We saw some examples of the grave problems associated with simulating a Hilbert space with a much smaller one. So we naturally ask: why are black holes special? Quantum and thermal systems have macroscopic observables which can be measured experimentally, and in a certain sense, we can find out the Hilbert space's correct dimension. Thus the possible kinematic overcounting of Hilbert space isn't realized in these systems. For the case of black holes, thermodynamic observables do point out that the Hilbert space is far smaller than what bulk semiclassical quantization indicates. As an example, we know that the thermodynamic entropy of a black hole should go as area. Further, from the CFT side we showed using various examples that spatially separated bulk interior states indeed have small inner products. These small overlaps between the dual bulk semiclassical excitations hint at how the Hilbert space of gravity embeds bulk states.

 We pose another question: Why we do not see an overcounting using effective field theory in empty AdS? Given a holographic CFT, the HKLL prescription wholly reconstructs the empty AdS bulk. As a result, the CFT description captures all the bulk excitations, and consequently, there is no question of any overcounting. The HKLL prescription readily reconstructs the exterior regions of eternal black holes as well. The only places where overcounting using effective field theory can arise are causally inaccessible regions from the boundary. The black hole interior is an example of such an inaccessible region, and reconstruction using state-dependent operators allows us to resolve apparent paradoxical situations, like the one we have treated here.

 In this regard, an important aspect that we have briefly touched upon here is the subject of quantum state transfers in semiclassical gravity. We showed via an example in \S \ref{genproperties} that overcounted Hilbert spaces can lead to quantum state transfers which are forbidden if orthogonal vectors span the Hilbert space. Since we have argued that effective field theory in the black hole interior leads to an overcounting of the Hilbert space, it will be useful to understand which forbidden semiclassical quantum state transfers are actually allowed in quantum gravity. We think this holds important implications for black holes in AdS and possibly in flat space as well.

 For the general reader not interested in details, we have thus answered an interesting puzzle: Can there exist a giant universe inside a big AdS black hole having a relatively small BH entropy, while an external observer is utterly oblivious to the universe's existence? The answer to this question is yes, provided that the Hilbert space of this universe is constructed from the small Hilbert space of the AdS black hole using small inner products. A significant number of humans, planets and stars can all be described using the overcounted Hilbert space, given that their backreaction on the black hole is extremely small (or in other words, their states belong to the small Hilbert space). Careful measurements of thermodynamic observables and state transfers in semiclassical gravity can lead to the conclusion that the Hilbert space of this universe is constructed from the small Hilbert space of the black hole itself. The initial state giving rise to this universe is a Euclidean state as mentioned in \S \ref{initialHH}, and the future of this universe is doomed as expected.

 \begin{acknowledgments}
I thank Suvrat Raju for collaboration in the initial stages of this work and also for extremely valuable suggestions and discussions regarding this work. I also thank Suvrat Raju, Akhil Sivakumar and Edward Witten for important comments on the initial draft. I am also grateful to Varun Dubey, Diksha Jain, Chandan Jana, R. Loganayagam, Alexandre Serantes, Pushkal Shrivastava and Akhil Sivakumar for various helpful discussions regarding this work. I acknowledge gratitude to the people of India for their steady and generous support to research in basic sciences.

\end{acknowledgments}

\appendix

\section{Review of construction of interior operators in the state-dependent formalism}
\label{StateDepReview}
In this section, we review the construction of state-dependent operators describing modes behind the horizon \cite{Papadodimas:2012aq, Papadodimas:2013jku, Papadodimas:2013wnh, Papadodimas:2015xma, Papadodimas:2015jra}. This construction is similar for both pure and eternal black holes. For eternal black holes, the CFT observables mean the right CFT's observables, which are our objects of interest.

We start with the black hole state $\ket{\psi}$ whose average energy is given by $E$. Firstly an algebra $\mathcal{A}$ is generated by \textit{simple operators} which are defined as operator polynomials of degree $n$ such that $n \ll \mathcal{N}$, where $\mathcal{N}$ is the central charge of CFT ($\mathcal{N} = N^2$). We will call this algebra the \textit{small algebra}. The small algebra is associated with the state $\ket{\psi}$ and does not include the Hamiltonian. We exclude the Hamiltonian because we do not want to include any annihilation operators in the algebra, and the Hamiltonian annihilates the state, i.e. $(H - E) \ket{\psi} = 0$. However, we want the algebra to be approximately closed under time evolution. We then construct and work in the \textit{small Hilbert space}, which is obtained by the action of these simple operators on the state $\ket{\psi}$.

\beq
\mathcal{H}_{\ket{\psi}} := \mathcal{A} \ket{\psi}
\eeq

We have thus laid out the basic framework in order to derive various axiomatic/algebraic QFT results. If the algebra $\mathcal{A}$ is considered a Von Neumann algebra, we can derive the Tomita-Takesaki theorem which constructs a commutant algebra $\mathcal{A}'$ for us. Note that this construction does not involve a doubling of the Hilbert space. We define the following antilinear map from $S: H_{\ket{\psi}} \to H_{\ket{\psi}}$ and $O \in \mathcal{A}$.

\begin{equation}
S \, O\ket{\psi} = O^{\dagger} \ket{\psi} \label{antilinear}
\end{equation}

We now decompose the operator $S$ as $S = \mathrm{J} \, \Delta^{1/2}$, where $\mathrm{J} $ is an anti-unitary operator and $\Delta$ is Hermitian. Consequently we have $S^{\dagger} S = \Delta$. The Tomita-Takesaki theorem says that there exists a commutant algebra $\mathcal{A}' \equiv \mathrm{J}  \, \mathcal{A} \, \mathrm{J} $, with the property that operators $\widetilde{O} \in \mathcal{A}'$ defined by $\widetilde{O} = \mathrm{J} \,  O \, \mathrm{J} $ commutes with all elements $O' \in \mathcal{A}$
\begin{equation}
\left[\widetilde{O}, O'\right] = 0.
\end{equation}

Since $\Delta$ is Hermitian we express it as $\Delta = \exp{-K}$, where $K$ is defined as the modular Hamiltonian for the algebras $\mathcal{A}$ and $\mathcal{A'}$ generating the Hilbert space $\mathcal{H}_{\ket{\psi}}$, and is expressed in terms of the antilinear operator $S$ as:
\beq
K = - \log{S^{\dagger} S}.
\eeq

Our job is to construct now the precise form of the modular Hamiltonian and the tilde operators. To construct these, we will apply the above construction to a system with Hamiltonian $H$ acting on the state $\psi$. As given in \cite{Papadodimas:2017qit}, the modular Hamiltonian up to the leading order in $N$ takes the form:
\beq
K = \beta (H - E) + \text{O} \,\lc \frac{1}{N} \rc
\eeq
where $E$ is the average energy of the state $\psi$ on which it is acting. Therefore to the leading order in $N$ one can give a precise form for the $\widetilde{O}$ operators. Using the definition given in equation \eqref{antilinear} with the definitions given by
$S^{\dagger} S = \Delta$ and $\widetilde{O} = \mathrm{J} \,  O \, \mathrm{J}$, we can write down the action of the $\widetilde{O}$ operators on the Hilbert space:
\beq
\widetilde{O}(\omega) \, O' \ket{\psi} = O' \, e^{-\beta \omega} \, O^{\dagger}(\omega) \ket{\psi} \quad \& \quad [H, \widetilde{O}(\omega)] \, O' \ksi = \omega\,  \widetilde{O}(\omega) \, O' \ksi.
\eeq

The state-dependent operators $\widetilde{O}$ describing the right moving modes in the interior are thus constructed in the above fashion. The commutant algebra allows us to impose causality and locality between the interior and the exterior operators. The role of modular operators is to push the excitations behind the horizon. The unique feature of this construction is that this follows naturally for any well-defined quantum field theory, provided the algebra of simple operators satisfies the requirements as mentioned above. 

\section{Explicit examples of overcounting in small vector spaces}
\label{lowdim}

In this appendix we explicitly demonstrate that there can be many more vectors than $n$, where $n$ is the dimensionality of the vector space if $\epsilon \neq 0$. We will demonstrate this using regular polyhedra. The study of overcounting using regular polyhedra serves as an easy way to develop our intuition for understanding overcounting by starting from small dimensions and gradually building up to higher dimensional examples. Note that our vector space defined over reals in contrast to the Hilbert spaces in quantum mechanics. We will denote the maximum number of vectors as a function of $\epsilon$ to be $\hat{m}_n (\epsilon)$ in the previous case. Now regular polyhedra are classified into three classes:

\begin{itemize}
\item 
\textit{Simplex:} This polyhedra is defined by the condition that the distance between any two vertices is the same. Examples are : equilateral triangle with $n =2$ and tetrahedron with $n =3$.  The vectors corresponding to neighbouring vertices have an inner product given by $\vec{p_i}. \vec{p_j} = -\frac{1}{n}$. Therefore we have $\hat{m}_n \lc -\frac{1}{n}\rc  = n+1$ number of vectors for the simplex.
\item 
\textit{Orthoplex:} These polyhedra are defined such that they have a vector each pointing towards each coordinate direction, suc that the inner product between the neighboring vertices is given by $\vec{p_i}. \vec{p_j} = 0$. Consequently a simplex has $ \hat{m}_n \lc 0\rc = 2n$ number of vectors.
\item 
\textit{Hypercube:} In the Cartesian coordinate system, these polyhedra have vertices situated at the coordinates $(\pm 1,\pm 1,\pm 1,\pm 1\dots)/\sqrt{n}$. Examples are : square with $n =2$ and cube  with $n =3$. The neighbouring vertices have inner products given by $\vec{p_i}. \vec{p_j} = 1-\frac{2}{n}$. Consequently a hypercube has $ \hat{m}_n \lc 1-\frac{2}{n}\rc = 2^n$ number of vectors.
\end{itemize}

 We will now compute the inner products for representatives of these above-mentioned classes of polyhedra. We will now give some examples in low dimensions below:

\begin{itemize}
\item 
\textit{Two dimensions:} In $n=2$ for a regular polygon,  the scalar product between position vectors of $m$ neighbouring vertices is given by $\vec{p_i}. \vec{p_j} = \cos \ls \frac{2\pi}{m}\rs$, where $\hat{m}_2 \lc \frac{2\pi}{m}\rc  = m$.
\item 
\textit{Three dimensions:} In $n=3$, the icosahedron has $\hat{m}_3 \lc \frac{1}{\sqrt{5}}\rc = 12$ vertices while the dodecahedron has $\hat{m}_3 \lc \frac{\sqrt{5}}{3}\rc = 20$ vertices. 
\item 
\textit{Four dimensions:} In $n=4$, we consider the 24-cell which has total number of vertices given by $\hat{m}_4 \lc \frac{1}{2}\rc = 24$. Similarly the 120-cell has $\hat{m}_4 \lc \frac{1+\sqrt{5}}{4}\rc = 120$ vertices while the 600-cell has $\hat{m}_4 \lc \frac{1+3\sqrt{5}}{8}\rc = 600$ vertices.

\end{itemize}

\subsection*{Overcounting in the limit $m \to \infty$?}

We will construct a situation where $m$ vectors are approximately equidistant on the sphere $S_{n-1}$ which has surface area of $\frac{2\pi^{n/2}}{\Gamma(\frac{n}{2})}$. Each unit vector has an exclusion zone given by $\frac{2\pi^{n/2}}{m\Gamma(\frac{n}{2})}$, where we cannot have any other vector. These exclusion zones have a radius $r$, with the volume of these  $n-1$ dimensional zones given by $\frac{\pi^{\frac{n-1}{2}}}{\Gamma(\frac{n+1}{2})}r^{n-1}$.  Therefore we can solve for the radius of the exclusion zone as done in the following equation:
\beq
r\approx\left(\frac{2\sqrt{\pi}\Gamma(\frac{n+1}{2})}{m\Gamma(\frac{n}{2})}\right)^{\frac{1}{n-1}}.
\eeq
 The distance between two neighbouring unit vectors is given by $d\approx2r$. Therefore the inner product between two neighbouring unit vectors can be easily computed and is found to be: 
 \beq
 \vec{p}_i\cdot\vec{p}_j\approx1-\frac12d^2=1-2\left(\frac{2\sqrt{\pi}\Gamma(\frac{n+1}{2})}{m\Gamma(\frac{n}{2})}\right)^{\frac{2}{n-1}}.
 \eeq
 
We can now solve the expression for inner products to obtain $m$. In the limit $\epsilon\to1$ with $0<1-\epsilon\ll1$, we obtain the following expression for $m$:

\beq
\hat{m}_n(\epsilon)\approx 2^{n/2}\frac{\sqrt{2\pi}\Gamma(\frac{n+1}{2})}{\Gamma(\frac{n}{2})}(1-\epsilon)^{-\frac{n-1}{2}}.
\eeq

 For $0< \epsilon \ll 1$, $m$ takes the values:
 
\beq
\hat{m}_n(\epsilon)\approx2n\left(\frac{2n}{n+1}\right)^{n\epsilon} \label{epsmall}
\eeq

This expression for $m$ is in agreement with our derivation of inner products between neighbouring vectors of the simplex and orthoplex, which are given by $\hat{m}_n\lc -\frac{1}{n}\rc = 1+n$ and $\hat{m}_n\lc 0\rc = 2n $ respectively. Equation \eqref{epsmall} is also in agreement with our derivation for the hypercube's case, up to a minor factor of $\frac{n}{2e}$.

\beq
\lim_{n \to \infty} \hat{m}_n\lc 1 - \frac{2}{n}\rc \to 2^n \times \frac{n}{2e}.
\eeq 

\section{Partition function of the $U(N)$ two-oscillator model}
\label{partitionfunction}
We outline the calculation of the partition function of two bosonic oscillators denoted by $A$ and $B$ transforming under the adjoint representation of $U(N)$ global group \cite{Aharony:2003sx, Sundborg:1999ue}. We hereby define $x = e^{-\beta}$.

Since we are interested only in the global group singlets, we calculate the partition function by summing over all Boltzmann weights multiplied by the number of group singlets at each Boltzmann weight. With $E_i$ being the energy of the bosonic modes, the partition function is given by

\begin{equation}
Z(x) = \sum_{n_1} \sum_{n_2}\, x^{n_1 E_1} x^{n_2 E_2}  \left(\text{All singlets } \text{sym}_{n_1}[\text{adj}]\times \text{sym}_{n_2}[\text{adj}] \right)  \label{partition1}
\end{equation}

Here the index $n_1$ goes over the $A$ oscillators and $n_2$ goes over $B$ oscillators. Note that here we have set the ground state energy of the oscillators to zero. We now return to the problem of counting the group singlets. A convenient way to count the number of group singlets is by using properties of characters, which are maps from the representation of the group to complex numbers defined as

\begin{equation}
\chi_{\mathbb{R}} : G \to \mathbb{C}
\end{equation}
satisfying the property that for $g \in G$,
\begin{equation}
\chi_{\mathbb{R}}(g) = \text{Tr}_{\mathbb{R}}(g).
\end{equation}

These satisfy the orthonormality relation, where  $[dg]$ is the Haar measure 

\begin{equation}
\int [dg] \, \chi^*_{\mathbb{R}_i}(g) \, \chi_{\mathbb{R}_j}(g) = \delta_{{R}_i{R}_j},
\end{equation}
chosen such that $\int [dg] = 1$. Since the character is the trace of the element, therefore $\chi_{\mathbb{R}_1 \times \mathbb{R}_2} = \chi_{\mathbb{R}_1} \, \chi_{\mathbb{R}_2}$. The number of irreps can thus be counted using the above relation as
\begin{equation}
n_{R_I} = \int [dg] \,  \chi^*_{\mathbb{R}_I}(g) \prod_j \chi_{\mathbb{R}_j}(g).
\end{equation}

We will now use this relation to count the number of group singlets. For a singlet representation, by definition we have $\chi_s = 1$. Therefore the number of singlets is given by 
\begin{equation}
\text{All singlets } = \int [dg] \, \prod_j \chi_{\mathbb{R}_j}(g) \label{singlet}
\end{equation}

We will now use \eqref{singlet} to rewrite the partition function in \eqref{partition1} in terms of the characters of $U(N)$:

\begin{equation}
Z(x) = \int [dU] \prod_{i =A,B} \, \sum_{n_i = 0}^{\infty} \,  x^{n_i E_i}  \, \chi_{\text{sym}^{n_i}}(U).
\end{equation}

We will utilize the following relation for the characters in order to simplify the partition function:

\begin{equation}
\sum_{n=0}^{\infty} t^n \,  \chi_{\text{sym}_n}(g) = \exp{\sum_{l=1}^{\infty}\frac{ t^l \,  \chi (g^l)}{l}} \label{chi1}
\end{equation}

We denote $z_A$ and $z_B$ as the single particle partition functions for the $A$ and the $B$ harmonic oscillators. Using \eqref{chi1}, and using the bosonic partition function $z(x) = z_A + z_B$ and $\chi_{\text{adj}}(U) = \text{Tr}(U) \, \text{Tr}(U^{\dagger})$ we get the partition function as

\begin{equation}
Z(x) = \int [dU]\, \exp\left[\sum_{k =1}^{\infty} \frac{z(x^k)}{k} \, \text{Tr}U^k\, \text{Tr}(U^{\dagger})^k\right].
\end{equation}

Rewriting this unitary matrix model in terms of the eigenvalues of the unitary matrix easily solves the model \cite{mehta2004random, Oxford}.  Denoting the eigenvalues of $U$ by $e^{i \alpha_i}$, we write the measure of the model as:
\begin{equation}
\int [dU] = \prod_i \int_{-\pi}^{\pi}[d \alpha_i] \, \prod_{i<j} \sin^2\frac{(\alpha_i - \alpha_j)}{2}
\end{equation}

Now the partition function is a function of the eigenvalues $\alpha_i$ and is given by
\begin{equation}
Z(x) = \int [d\alpha_i] \exp{-\sum_{i \neq j}V(\alpha_i - \alpha_j)}
\end{equation}
where the potential is given by
\begin{equation}
\begin{split}
V(\theta) &= - \log\abs{\sin\frac{\theta}{2}} - \sum_{k=1}^{\infty} \frac{z(x^k)}{k} \cos k\theta\\
\end{split} \label{potential}
\end{equation}

Equation \eqref{potential} is reminiscent of the Coulomb potential for charges on a sphere where the term coming from the measure is repulsive interaction between the like charges, and the other term is attractive interaction due to electric field. The dynamics are similar to the partition function of the Gross-Witten-Wadia model \cite{Gross:1980he, WADIA1980403, wadia2012study}, and has a third-order phase transition in the $N \to \infty$ limit and has a free energy of O $(N^2)$. The average energy is $\text{O}\, (N^2)$, and as a consequence, the entropy is also of the same order. Therefore we can see that the Hilbert space's dimensionality is $\text{O} \,(\exp N^2)$.

\subsection*{Entropy, average energy and dimensionality of the Hilbert space for oscillators on $S_{3} \times \mathbb{R}$}

 We now proceed to calculate the Hagedorn temperature of the CFT, which is essential because we want to describe black holes, and going above the Hagedorn temperature is the regime where we have black holes in the bulk. We will then calculate the entropy, the average energy and the dimensionality of the Hilbert space. 

\subsubsection*{The single particle partition functions}

We evaluate the single-particle partition functions for the CFT living on $S^{3} \times \mathbb{R}$. With $\lambda \approx 0$ our bosonic harmonic oscillators $A$ and $B$ obey the equation of motion given by 

\beq
\lc -\del^2 + a^{-2} \rc A_{ij} =0 \quad \& \quad \lc -\del^2 + a^{-2} \rc B_{ij} =0
\eeq

In four dimensions we utilize the conformal map from $S^{3} \times \mathbb{R} \to \mathbb{R}^4$ to write the partition function $Z = \sum_{E_i} e^{-\beta E_i}$ in the form $Z = \sum_{\Delta} e^{-\beta \Delta}$ where $\Delta$ is the scaling dimension. The scaling dimension goes over all the local operators in the theory which are generated by repeated applications of the derivatives $\del_{\mu}$ on the fields, i.e. $\lc A, \del_{\nu} A, \del_{\mu}\del_{\nu} A \dots\rc$ and similarly for the $B$ field modulo the equation of motion. Since $\ls \del \rs =1$ a single derivative gives rise to a factor in:
\beq
x^0 + x^1 + x^2 +x^3+ \dots = \frac{1}{1-x}
\eeq

Four such derivatives will give rise to $\frac{1}{(1-x)^4}$. The mass dimension of the matrix oscillators is $[A], [B] = 1$. Without incorporating the equation of motion, the naive partition function constructed using all local operators arising in the matrix model is given by

\beq
\frac{x}{(1-x)^4}
\eeq
We now need to take the modulus by the equation of motion. Notice that the equation of motion imposes a condition on any local operator $O$ of the theory:

\beq
\lc -\del^2 + a^{-2} \rc O = 0.
\eeq

The factor $\lc -\del^2 + a^{-2} \rc$ has a mass dimension $x^2$ which we need to subtract off. The CFT partition function for oscillators $A$ and $B$ upon this subtraction is therefore given by
\beq
z_A(x)=z_B(x) = \frac{x-x^3}{(1-x)^4} = \frac{x+x^2}{(1-x)^3}.
\eeq

\subsubsection*{The Hagedorn transition}
We calculate the entropy in this subsection and deduce the dimensionality of the Hilbert space from it. First of all we set $m_A = m_B =1$ for our convenience. The entropy from the partition function with $k_B =1$ is given by
\beq
S =-\frac{\del F}{\del T} = \frac{\del \lc T \log Z\rc}{\del T} = \log Z + \frac{T}{Z} \frac{\del Z}{\del T}.
\eeq

We can write the potential in \eqref{potential} as

\beq
V(\theta) = \log 2 + \sum_{k=1}^{k =\infty} \frac{1}{k} \lc 1- \frac{2( x+x^2)}{(1-x)^3}\rc  \, \cos  k \theta.
\eeq

In the low-temperature phase, it follows from \eqref{potential} that the attractive second term goes to zero as $T \to 0$. Therefore at low temperature, the repulsive interactions dominate. The way to solve this matrix model in \eqref{potential} is to introduce the eigenvalue density and solve using the mean-field theory approximation. The level density of the eigenvalues is spread uniformly over the circle as $T \to 0$. As we increase the temperature, this distribution becomes more and more non-uniform, and the phase ceases to be stable when the terms in the potential turn negative. The condition for the stability of the potential then becomes

\beq
\frac{2( x+x^2)}{(1-x)^3} < 1.
\eeq 

Since $0<x<1$ and $x$ monotonically increases with temperature in this regime, the leading $k=1$ order well approximates the above condition since it gives the strongest contribution. The temperature at which this phase becomes unstable is the Hagedorn temperature and is given by 

\beq
\frac{2( x+x^2)}{(1-x)^3} = 1.
\eeq

 Solving this the Hagedorn temperature is given by $k_b T_H = 0.634484$.

\subsubsection*{Evaluation of the partition function for small $\mathcal{N}$}

We use the single-particle bosonic partition functions to expand the complete partition function up to the first two powers in cosines, as the remaining terms fall off quite rapidly and therefore have negligible contributions. We will demonstrate this for the $N=2$ case and will treat higher $N$ similarly. The partition function for $N=2$ is given by
 
 \beq
Z(x) = \int_{-\pi}^{\pi} \int_{-\pi}^{\pi} d\alpha_1 d\alpha_2 \, \sin^2 \lc \frac{\alpha_2 - \alpha_1}{2}\rc \exp \ls \sum_{k=1}^{\infty} \frac{4\cos k\lc \alpha_2 - \alpha_1\rc}{k} \frac{x+x^2}{(1-x)^3}\rs.
\eeq

Upon the expansion to the first two orders, we obtain the partition function to be
 
 \beq
Z(x) = \int_{-\pi}^{\pi} \int_{-\pi}^{\pi} d\alpha_1 d\alpha_2 \, \sin^2 \lc \frac{\alpha_2 - \alpha_1}{2}\rc \exp \ls  \frac{4 (x+x^2) \cos \lc \alpha_2 - \alpha_1\rc }{(1 - x)^3} +   \frac{2(x^2+x^4)\cos \ls 2\lc \alpha_2 - \alpha_1\rc \rs}{(1 - x^2)^3}\rs. \label{N=2}
\eeq

We evaluate the partition function numerically for $N=2$ using equation \eqref{N=2} at temperature $T = 0.64$, which is just above the Hagedorn temperature. Similarly, we can explicitly write down the first two terms in the partition function, which are the leading contributions and numerically integrate them for $N =3, 4,5$ at $T = 0.64$. We use these to derive the numerical values of entropy, average energy and use the entropy to calculate the dimensionality of the Hilbert space. These values are given in Table \ref{table}.

\section{Maximum volume slices for the $AdS$ black hole}
\label{Appendix 1}

In this appendix we will maximize the volume of the nice slices of $AdS_{d+1}$ black holes whose endpoints are at $(u_0,0)$ on the left horizon and $(0,v_0)$ on the right horizon.  Using the isometry of $AdS$, we will set $u_0 = v_0$. The following calculation holds for both the one-sided black hole and the eternal black hole. The metric in the Kruskal coordinates for the eternal black hole is given by
\begin{equation}
ds^2 = -\frac{4f(r)}{f'(r_h)^2} \, e^{-f'(r_h) r^*} du_k \, dv_k + r^2 d\Omega^2_{d-1}, 
\end{equation}
where $f(r) = r^2 + 1 - \frac{C}{r^{d-2}}$, $r_h$ is the black hole horizon and $r^*$ is the tortoise coordinate. The Kruskal coordinates are denoted with a subscript $k$ in order to avoid potential confusion with Eddington-Finkelstein coordinate $v$ which we will be using later on.

\subsection*{The variational problem}
In this subsection, we utilize the method given in \cite{Stanford:2014jda, Carmi:2017jqz} to compute the volumes of maximal slices. First, we will define a conserved quantity $E$ and write the maximum volume in terms of it. Afterwards, we will fix $E$ in terms of the Kruskal coordinate $u_0$.

Note that the method used to calculate the volume-maximizing slices is not restricted to AdS black holes. As an example, we will calculate the maximum volume for slices in the interior of a $2+1$ dimensional AdS black brane.

\subsubsection*{Expression for the maximum volume in terms of the conserved quantity $E$}

We write the black hole metric in infalling Eddington-Finkelstein coordinates where we will be using $v = t+ r^*$. The method used to compute the volume does not depend on a specific $f(r)$. In these coordinates, the metric takes the form:
\begin{equation}
ds^2 = -f(r)\, dv^2 + 2 dr\, dv + r^2 d\Omega^2_{d-1},
\end{equation}
where again $f(r) = r^2 + 1 - \frac{C}{r^{d-2}}$. We assume that the nice slice of maximum volume has the same symmetry as that of the $(d-1)$ sphere. Extremizing \eqref{eq1} gives the maximum volume, where a dot denotes the derivative with respect to $\sigma$, which is a parameter characterizing the nice slices. Here $V_{d-1}$ is the volume of the $(d-1)$ spherical ball.

\begin{equation}
V = V_{d-1}\int d\sigma \, r^{d-1} \left( -f(r) \, \dot{v}^2 + 2\dot{r} \, \dot{v}\right)^{\frac{1}{2}}   \label{eq1}
\end{equation}

Extremizing the above equation follows the same procedure for extremizing action, with the integrand playing the role of a Lagrangian. Since the Lagrangian does not depend on $v$, therefore we have a conserved quantity $E = - \frac{\partial L}{\partial \dot{v}}$.

The volume here \eqref{eq1} is reparametrization invariant as it does not depend on the choice of $\sigma$. We will fix the parametrization as follows:
\begin{equation}
r^{d-1}\left( -f(r) \, \dot{v}^2 + 2\dot{r} \, \dot{v}\right)^{\frac{1}{2}} - 1 = 0.
\end{equation}

We can now write down equations determining $r(\sigma)$ and $v(\sigma)$ using the fixed parametrization and the expression for $E$, which allows us to write down $r$ and $v$ as coupled differential equations in terms of E. 

\begin{equation}
E - r^{2(d-1)}\left[ f(r)\dot{v} - \dot{r}\right] = 0 \label{radius1}
\end{equation}
\begin{equation}
 r^{2(d-1)}\dot{r}^2 - f(r) - r^{-2(d-1)}E^2 = 0 \label{radius2}.
 \end{equation}
 
 Eliminating $\dot{v}$ using the above equation, the expression for the maximum volume takes the following form:
 \begin{equation}
 V = 2 V_{d-1} \int_{r_{min}}^{r_{h}} \frac{dr}{\dot{r}} = 2 V_{d-1} \int_{r_{min}}^{r_{h}} dr \,  \frac{r^{2(d-1)}}{\sqrt{f(r) r^{2(d-1)}+ E^2}} \label{vol},
 \end{equation}
 where $r$  in the integral goes to a minimum value of $r_{min}$ which is determined by substituting $\dot{r} = 0$ in \eqref{radius2}. 
 
 \begin{equation}
 f(r_{min}) \, r^{2(d-1)}_{min} + E^2 =0. \label{rmin}
 \end{equation}

\subsubsection*{Fixing $E$ in terms of Kruskal coordinate $u_0$}
 Here we fix $E$ in terms of $u_0$. We see that $E$ is negative since at $r = r_{min}$ as we have $\dot{r}|_{r= r_{min}} = 0$ and $\dot{v}|_{r = r_{min}} > 0$.   From the definition of the coordinate $v$, and using equations \eqref{radius1} and \eqref{radius2} we get
 \begin{equation}
 t_r + r_{h}^* - r^*(r_{min}) = \int^{v_{r_h}}_{v_{min}} dv = \int_{r_{min}}^{r_{h}} dr \, \Bigg[ \frac{E}{f(r)\sqrt{f(r)r^{2(d-1)}+E^2}}+\frac{1}{f(r)}\Bigg] \label{v}
 \end{equation}

Inside the horizon, the Kruskal coordinate $u_k$ is related to $r^*$ and $t$ as $u = e^{\frac{f'(r_h)}{2}(r^* -t)}$. Expressed in Kruskal coordinates \eqref{v} is given by:
\begin{equation}
\log(u_0) = \log u_{min} + \frac{f'(r_h)}{2} \int_{r_{min}}^{r_{h}} dr \, \Bigg[ \frac{E}{f(r)\sqrt{f(r)r^{2(d-1)}+E^2}}+\frac{1}{f(r)}\Bigg] \label{Evsu}
\end{equation}

\subsection*{Analytic expression for the volume growth of maximum volume surfaces in terms of Kruskal coordinate $u_0$}

We derive an analytic expression relating the volume growth of the maximum volume surfaces in terms of the Kruskal coordinate $u_0$. Notice that the integrands in \eqref{vol} and \eqref{Evsu} are regular and don't blow up at $r_h$. The integrands are denoted below by $V'$ and $I'$ respectively.

\begin{equation}
\lim_{r \to r_h} \text{V}' = \frac{V_{d-1}r_h^{2(d-1)}}{|E|}
\end{equation} 

\begin{equation}
\lim_{r \to r_h} \text{I}' = \lim_{r \to r_h} \Bigg[\frac{E}{(r^2 -r_h^2)\sqrt{(r^2-r_h^2)r^2 +E^2}} + \frac{1}{f(r)} \Bigg] = \frac{1}{4} \frac{f'(r_h)r_h^{2(d-1)}}{E^2}
\end{equation}

At $r_{min}$ also both these integrands encounter a similar logarithmic blow-up.

\begin{equation}
\lim_{r \to r_{min}} \text{V}' = \frac{V_{d-1}r_{min}^{2(d-1)}}{\sqrt{E^2 + r^{2(d-1)} f(r)}}
\end{equation} 

\begin{equation}
\lim_{r \to r_{min}} \text{I}' = \lim_{r \to r_{min}} \Bigg[\frac{E}{(r^2 -r_h^2)\sqrt{(r^2-r_h^2)r^2 +E^2}} + \frac{1}{f(r)} \Bigg] = -\frac{f'(r_h)|E|}{2 f(r_{min})r_{min}^{2(d-1)}} \frac{r_{min}^{2(d-1)}}{\sqrt{E^2 + r^{2(d-1)} f(r)}}
\end{equation}

Since these integrands have similar blow up, one can relate the volume with $u_0$ as follows using \eqref{rmin} and the definition of Hawking temperature $\beta$,
\begin{equation}
V = \frac{\beta A(r_{min})}{2 \pi}  \log{u_0} + \text{O}\,(1) \label{Vanalytic}
\end{equation}

Here $\text{O}\,(1)$ is a subleading quantity which does not grow with $u_0$.

\subsection*{An example: The 2+1 black brane}

As an example of this above volume maximization we will take a look at $2+1$ dimensional branes. This provides us an oppurtunity to study the late-time behaviour of $E$, thereby allowing us to understand how $E$ characterizes the slice. The general $d+1$ black brane metric is given by:
\begin{equation}
ds^2 = -f(r)\, dv^2 + 2 dr\, dv +  r^2 dx^2_{d-1},
\end{equation}
where $f(r) = r^2\left(1 - \lc \frac{r_h}{r}\rc^d\right)$. The inverse temperature for the black brane is given by $\beta = \frac{4\pi}{f'(r_h)} = \frac{4\pi}{d r_h}$. The expression for $u_{min}$ takes the form
\begin{equation}
u_{min} = e^{\frac{d r_h}{2} r^*(r_{min})}, 
\end{equation}
where $r_{min}$ is calculated from \eqref{rmin} for $d=2$ to be 
\begin{equation}
r_{min} = \sqrt{\frac{r_h^2 + \sqrt{r_h^4 - 4E^2}}{2}}. \label{rmin}
\end{equation}

Note that \eqref{rmin} has 3 more roots, two of which are dropped because they are negative. The third root $r = \sqrt{\frac{r_h^2 - \sqrt{r_h^4 - 4E^2}}{2}}$ is dropped as the integrand becomes imaginary once the lower limit of integration goes below $r_{min}= \sqrt{\frac{r_h^2 + \sqrt{r_h^4 - 4E^2}}{2}}$. The volume and the relation of $E$ with $u_0$ for $d=2$ are respectively given by
\begin{equation}
V = \int_{r_{min}}^{r_h} dr \, \frac{r^2}{\sqrt{(r^2 - r_h^2)r^2 + E^2}} \label{3V}
\end{equation}
and
\begin{equation}
\log{u_0}=\log{u_{min}} + r_h\int_{r_{min}}^{r_h} dr \, \Bigg[\frac{E}{(r^2 -r_h^2)\sqrt{(r^2-r_h^2)r^2 +E^2}} + \frac{1}{f(r)} \Bigg] \label{3uE}
\end{equation}

\subsubsection*{Late time behaviour of $r_{min}$ and $E$}
The indefinite integrals indicate that the volume tends to infinity as $E^2 \to \frac{r_h^4}{4}$. We took into account the largest root as $r_{min}$ while solving the minimization equation $E^2 + r_{min}^2 f(r_{min}) = 0$. $|E|$ characterizes the nice slice and it increases monotonically with Kruskal time. Therefore at late times $r_{m} = \lim_{(u_0 \to \infty)} \, r_{min}$ is an extremum of $r^2 f(r)$, which translates to $[r^2 f(r)]' = 0$. By definition $r_m$ is also a root of $E^2 + r^2 f(r) = 0$. Therefore we see at late times,
\begin{equation}
r_m = \frac{r_h}{\sqrt{2}}, \label{rm}
\end{equation}
while late times $E$ is related to $r_h$ by
\begin{equation}
E^2 = \frac{r_h^4}{4}. \label{E}
\end{equation}

The volume is given by \eqref{Vanalytic}, where we substitute $r_{min}$ from \eqref{rmin}.

\section{Technique used to accommodate vectors on the unit sphere in Hilbert space}
\label{technique}

\begin{figure}

\centering
\includegraphics[width=.49\textwidth]{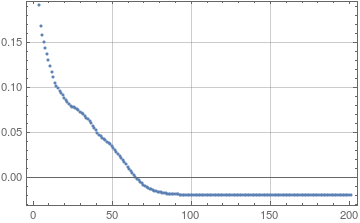}\hfill
\includegraphics[width=.49\textwidth]{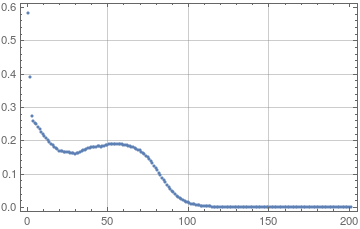}\hfill
\includegraphics[width=.49\textwidth]{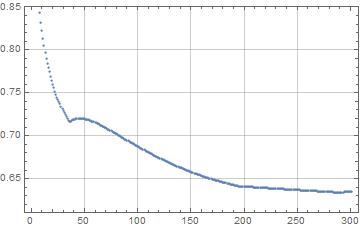}

\caption{We test the pushing technique for simplex in the left figure, orthoplex on the right and hypercube in the bottom figure. We see that the inner products converge to their actual value after around 100 iterations for the left and the right figures where we set the parameters $\alpha =100, \, \beta =4$. In comparison, inner products converge to their actual value after 200-220 iterations for the bottom figure which we perform with $\alpha =1$ and  $\beta=1$. }
\label{fig:figure3}

\end{figure}

In order to demonstrate overcounting, we construct a larger number of vectors on a sphere than its dimension such that they are almost equally separated from each other. This procedure is similar to the one we used to derive the worst-case overcounting formula.

In order to do this we define an energy function with a positive energy cost, i.e. a repelling force if vectors are too close to each other. The energy is minimized when the points are evenly distributed. We implement this numerically by pushing vectors away from each other. For vector $\vec{v_i}$ where $i \neq j$,

\beq
\vec{v_i} \to \vec{v_i} + \alpha \frac{\left(\vec{v_i} - \vec{v_j}\right)}{\left|\vec{v_i} - \vec{v_j}\right|^{2\beta}}
\eeq

This map drives a vector away from the nearest vectors and brings it closer to other distant vectors. The action pushes the vector out of the sphere, and we compensate this by normalizing the vector to bring them back on the sphere. $\alpha$ and $\beta$ are parameters which make the pushing action more or less. As a result, the pushing action separates the vectors until they come close to equilibrium and are almost equally separated from each other.

As a test, we check our program for the inner products of simplex, orthoplex and hypercube whose inner product we have already computed in Appendix \ref{lowdim}. We push these vectors $200,200, 300$ times respectively with $\alpha = 100, 100, 1$ for these three cases and $\beta = 4,4,1$, and find that there is convergence to the theoretically estimated dot product after approximately 120 iterations in each case. Figure \ref{fig:figure3} demonstrates the convergence of the maximum inner products to their theoretical values.

Figure \ref{fig:figure3} gives a nice description of what the method does. Stronger values of $\alpha$ and $\beta$ means a stronger repelling force from nearby vectors and hence more energy cost. Hence for $\alpha = 100, \beta =4$ we require about 120 iterations for the 300 points to converge. However, for a much lower value of $\alpha=1, \beta =1$ the repelling force is not that strong, and hence the energy cost is not great. Hence it takes many iterations for the inner products to converge to their theoretical value. 

\bibliographystyle{JHEP}
\bibliography{citation.bib}

\providecommand{\href}[2]{#2}\begingroup\raggedright\begin{thebibliography}{10}

\bibitem{Bekenstein-bhae}
J.~D. Bekenstein, {\it Black holes and entropy},  {\em Phys. Rev. D} {\bf 7}
  (Apr, 1973) 2333--2346.

\bibitem{Hawking-particle-creation}
S.~Hawking, {\it {Particle Creation by Black Holes}},  {\em Commun. Math.
  Phys.} {\bf 43} (1975) 199--220.

\bibitem{Maldacena:1997re}
J.~M. Maldacena, {\it {The Large N limit of superconformal field theories and
  supergravity}},  {\em Int. J. Theor. Phys.} {\bf 38} (1999) 1113--1133,
  [\href{http://arxiv.org/abs/hep-th/9711200}{{\tt hep-th/9711200}}].

\bibitem{Witten-ads-and-holography}
E.~Witten, {\it {Anti-de Sitter space and holography}},  {\em Adv. Theor. Math.
  Phys.} {\bf 2} (1998) 253--291,
  [\href{http://arxiv.org/abs/hep-th/9802150}{{\tt hep-th/9802150}}].

\bibitem{Gubser:1998bc}
S.~Gubser, I.~R. Klebanov, and A.~M. Polyakov, {\it {Gauge theory correlators
  from noncritical string theory}},  {\em Phys. Lett. B} {\bf 428} (1998)
  105--114, [\href{http://arxiv.org/abs/hep-th/9802109}{{\tt hep-th/9802109}}].

\bibitem{Hawking:1974sw}
S.~Hawking, {\it {Particle Creation by Black Holes}},  {\em Commun. Math.
  Phys.} {\bf 43} (1975) 199--220. [Erratum: Commun.Math.Phys. 46, 206 (1976)].

\bibitem{Mathur:2009hf}
S.~D. Mathur, {\it {The Information paradox: A Pedagogical introduction}},
  {\em Class. Quant. Grav.} {\bf 26} (2009) 224001,
  [\href{http://arxiv.org/abs/0909.1038}{{\tt arXiv:0909.1038}}].

\bibitem{Mathur:2012np}
S.~D. Mathur, {\it {The information paradox: conflicts and resolutions}},  {\em
  Pramana} {\bf 79} (2012) 1059--1073,
  [\href{http://arxiv.org/abs/1201.2079}{{\tt arXiv:1201.2079}}].

\bibitem{Susskind:1993if}
L.~Susskind, L.~Thorlacius, and J.~Uglum, {\it {The Stretched horizon and black
  hole complementarity}},  {\em Phys. Rev. D} {\bf 48} (1993) 3743--3761,
  [\href{http://arxiv.org/abs/hep-th/9306069}{{\tt hep-th/9306069}}].

\bibitem{Susskind:1993mu}
L.~Susskind and L.~Thorlacius, {\it {Gedanken experiments involving black
  holes}},  {\em Phys. Rev. D} {\bf 49} (1994) 966--974,
  [\href{http://arxiv.org/abs/hep-th/9308100}{{\tt hep-th/9308100}}].

\bibitem{Page:1993wv}
D.~N. Page, {\it {Information in black hole radiation}},  {\em Phys. Rev.
  Lett.} {\bf 71} (1993) 3743--3746,
  [\href{http://arxiv.org/abs/hep-th/9306083}{{\tt hep-th/9306083}}].

\bibitem{Strominger:1996sh}
A.~Strominger and C.~Vafa, {\it {Microscopic origin of the Bekenstein-Hawking
  entropy}},  {\em Phys. Lett. B} {\bf 379} (1996) 99--104,
  [\href{http://arxiv.org/abs/hep-th/9601029}{{\tt hep-th/9601029}}].

\bibitem{VanRaamsdonk:2010pw}
M.~Van~Raamsdonk, {\it {Building up spacetime with quantum entanglement}},
  {\em Gen. Rel. Grav.} {\bf 42} (2010) 2323--2329,
  [\href{http://arxiv.org/abs/1005.3035}{{\tt arXiv:1005.3035}}].

\bibitem{Hayden:2007cs}
P.~Hayden and J.~Preskill, {\it {Black holes as mirrors: Quantum information in
  random subsystems}},  {\em JHEP} {\bf 09} (2007) 120,
  [\href{http://arxiv.org/abs/0708.4025}{{\tt arXiv:0708.4025}}].

\bibitem{Almheiri:2013hfa}
A.~Almheiri, D.~Marolf, J.~Polchinski, D.~Stanford, and J.~Sully, {\it {An
  Apologia for Firewalls}},  {\em JHEP} {\bf 09} (2013) 018,
  [\href{http://arxiv.org/abs/1304.6483}{{\tt arXiv:1304.6483}}].

\bibitem{Verlinde:2012cy}
E.~Verlinde and H.~Verlinde, {\it {Black Hole Entanglement and Quantum Error
  Correction}},  {\em JHEP} {\bf 10} (2013) 107,
  [\href{http://arxiv.org/abs/1211.6913}{{\tt arXiv:1211.6913}}].

\bibitem{Shenker:2013pqa}
S.~H. Shenker and D.~Stanford, {\it {Black holes and the butterfly effect}},
  {\em JHEP} {\bf 03} (2014) 067, [\href{http://arxiv.org/abs/1306.0622}{{\tt
  arXiv:1306.0622}}].

\bibitem{Maldacena:2013xja}
J.~Maldacena and L.~Susskind, {\it {Cool horizons for entangled black holes}},
  {\em Fortsch. Phys.} {\bf 61} (2013) 781--811,
  [\href{http://arxiv.org/abs/1306.0533}{{\tt arXiv:1306.0533}}].

\bibitem{Jafferis:2017tiu}
D.~L. Jafferis, {\it {Bulk reconstruction and the Hartle-Hawking
  wavefunction}},  \href{http://arxiv.org/abs/1703.01519}{{\tt
  arXiv:1703.01519}}.

\bibitem{Penington:2019npb}
G.~Penington, {\it {Entanglement Wedge Reconstruction and the Information
  Paradox}},  \href{http://arxiv.org/abs/1905.08255}{{\tt arXiv:1905.08255}}.

\bibitem{Almheiri:2019hni}
A.~Almheiri, R.~Mahajan, J.~Maldacena, and Y.~Zhao, {\it {The Page curve of
  Hawking radiation from semiclassical geometry}},  {\em JHEP} {\bf 03} (2020)
  149, [\href{http://arxiv.org/abs/1908.10996}{{\tt arXiv:1908.10996}}].

\bibitem{Almheiri:2019qdq}
A.~Almheiri, T.~Hartman, J.~Maldacena, E.~Shaghoulian, and A.~Tajdini, {\it
  {Replica Wormholes and the Entropy of Hawking Radiation}},  {\em JHEP} {\bf
  05} (2020) 013, [\href{http://arxiv.org/abs/1911.12333}{{\tt
  arXiv:1911.12333}}].

\bibitem{Penington:2019kki}
G.~Penington, S.~H. Shenker, D.~Stanford, and Z.~Yang, {\it {Replica wormholes
  and the black hole interior}},  \href{http://arxiv.org/abs/1911.11977}{{\tt
  arXiv:1911.11977}}.

\bibitem{Papadodimas:2012aq}
K.~Papadodimas and S.~Raju, {\it {An Infalling Observer in AdS/CFT}},  {\em
  JHEP} {\bf 10} (2013) 212, [\href{http://arxiv.org/abs/1211.6767}{{\tt
  arXiv:1211.6767}}].

\bibitem{Papadodimas:2013jku}
K.~Papadodimas and S.~Raju, {\it {State-Dependent Bulk-Boundary Maps and Black
  Hole Complementarity}},  {\em Phys. Rev. D} {\bf 89} (2014), no.~8 086010,
  [\href{http://arxiv.org/abs/1310.6335}{{\tt arXiv:1310.6335}}].

\bibitem{Papadodimas:2013wnh}
K.~Papadodimas and S.~Raju, {\it {Black Hole Interior in the Holographic
  Correspondence and the Information Paradox}},  {\em Phys. Rev. Lett.} {\bf
  112} (2014), no.~5 051301, [\href{http://arxiv.org/abs/1310.6334}{{\tt
  arXiv:1310.6334}}].

\bibitem{Papadodimas:2015xma}
K.~Papadodimas and S.~Raju, {\it {Local Operators in the Eternal Black Hole}},
  {\em Phys. Rev. Lett.} {\bf 115} (2015), no.~21 211601,
  [\href{http://arxiv.org/abs/1502.06692}{{\tt arXiv:1502.06692}}].

\bibitem{Papadodimas:2015jra}
K.~Papadodimas and S.~Raju, {\it {Remarks on the necessity and implications of
  state-dependence in the black hole interior}},  {\em Phys. Rev. D} {\bf 93}
  (2016), no.~8 084049, [\href{http://arxiv.org/abs/1503.08825}{{\tt
  arXiv:1503.08825}}].

\bibitem{Wheeler}
J.~Wheeler, {\it {in Relativity, Groups and Fields, edited by B.S. DeWitt and
  C.M. DeWitt}},  {\em Gordon and Breach, New York} (1964).

\bibitem{Maldacena:2001kr}
J.~M. Maldacena, {\it {Eternal black holes in anti-de Sitter}},  {\em JHEP}
  {\bf 04} (2003) 021, [\href{http://arxiv.org/abs/hep-th/0106112}{{\tt
  hep-th/0106112}}].

\bibitem{Bohigas}
O.~Bohigas, M.~J. Giannoni, and C.~Schmit, {\it Characterization of chaotic
  quantum spectra and universality of level fluctuation laws},  {\em Phys. Rev.
  Lett.} {\bf 52} (Jan, 1984) 1--4.

\bibitem{mehta2004random}
M.~Mehta, {\em Random Matrices}.
\newblock ISSN. Elsevier Science, 2004.

\bibitem{doi:10.1063/1.1703773}
F.~J. Dyson, {\it Statistical theory of the energy levels of complex systems.
  i},  {\em Journal of Mathematical Physics} {\bf 3} (1962), no.~1 140--156,
  [\href{http://arxiv.org/abs/https://doi.org/10.1063/1.1703773}{{\tt
  https://doi.org/10.1063/1.1703773}}].

\bibitem{doi:10.1063/1.1703775}
F.~J. Dyson, {\it Statistical theory of the energy levels of complex systems.
  iii},  {\em Journal of Mathematical Physics} {\bf 3} (1962), no.~1 166--175,
  [\href{http://arxiv.org/abs/https://doi.org/10.1063/1.1703775}{{\tt
  https://doi.org/10.1063/1.1703775}}].

\bibitem{MEHTA1960395}
M.~Mehta, {\it On the statistical properties of the level-spacings in nuclear
  spectra},  {\em Nuclear Physics} {\bf 18} (1960) 395 -- 419.

\bibitem{Sekino:2008he}
Y.~Sekino and L.~Susskind, {\it {Fast Scramblers}},  {\em JHEP} {\bf 10} (2008)
  065, [\href{http://arxiv.org/abs/0808.2096}{{\tt arXiv:0808.2096}}].

\bibitem{Lashkari:2011yi}
N.~Lashkari, D.~Stanford, M.~Hastings, T.~Osborne, and P.~Hayden, {\it {Towards
  the Fast Scrambling Conjecture}},  {\em JHEP} {\bf 04} (2013) 022,
  [\href{http://arxiv.org/abs/1111.6580}{{\tt arXiv:1111.6580}}].

\bibitem{Maldacena:2015waa}
J.~Maldacena, S.~H. Shenker, and D.~Stanford, {\it {A bound on chaos}},  {\em
  JHEP} {\bf 08} (2016) 106, [\href{http://arxiv.org/abs/1503.01409}{{\tt
  arXiv:1503.01409}}].

\bibitem{Cotler:2016fpe}
J.~S. Cotler, G.~Gur-Ari, M.~Hanada, J.~Polchinski, P.~Saad, S.~H. Shenker,
  D.~Stanford, A.~Streicher, and M.~Tezuka, {\it {Black Holes and Random
  Matrices}},  {\em JHEP} {\bf 05} (2017) 118,
  [\href{http://arxiv.org/abs/1611.04650}{{\tt arXiv:1611.04650}}]. [Erratum:
  JHEP 09, 002 (2018)].

\bibitem{PhysRevLett.75.902}
A.~V. Andreev and B.~L. Altshuler, {\it Spectral statistics beyond random
  matrix theory},  {\em Phys. Rev. Lett.} {\bf 75} (Jul, 1995) 902--905.

\bibitem{PhysRevE.55.4067}
E.~Br\'ezin and S.~Hikami, {\it Spectral form factor in a random matrix
  theory},  {\em Phys. Rev. E} {\bf 55} (Apr, 1997) 4067--4083.

\bibitem{PhysRevE.56.264}
E.~Br\'ezin and S.~Hikami, {\it Extension of level-spacing universality},  {\em
  Phys. Rev. E} {\bf 56} (Jul, 1997) 264--269.

\bibitem{Marolf-bog}
D.~Marolf, {\it {Black Holes, AdS, and CFTs}},  {\em Gen. Rel. Grav.} {\bf 41}
  (2009) 903--917, [\href{http://arxiv.org/abs/0810.4886}{{\tt
  arXiv:0810.4886}}].

\bibitem{Hsu:2009kv}
S.~D. Hsu and D.~Reeb, {\it {Monsters, black holes and the statistical
  mechanics of gravity}},  {\em Mod. Phys. Lett. A} {\bf 24} (2009) 1875--1887,
  [\href{http://arxiv.org/abs/0908.1265}{{\tt arXiv:0908.1265}}].

\bibitem{Freivogel:2005qh}
B.~Freivogel, V.~E. Hubeny, A.~Maloney, R.~C. Myers, M.~Rangamani, and
  S.~Shenker, {\it {Inflation in AdS/CFT}},  {\em JHEP} {\bf 03} (2006) 007,
  [\href{http://arxiv.org/abs/hep-th/0510046}{{\tt hep-th/0510046}}].

\bibitem{Fu:2019oyc}
Z.~Fu and D.~Marolf, {\it {Bag-of-gold spacetimes, Euclidean wormholes, and
  inflation from domain walls in AdS/CFT}},  {\em JHEP} {\bf 11} (2019) 040,
  [\href{http://arxiv.org/abs/1909.02505}{{\tt arXiv:1909.02505}}].

\bibitem{Ong:2013mba}
Y.~C. Ong and P.~Chen, {\it {The Fate of Monsters in Anti-de Sitter
  Spacetime}},  {\em JHEP} {\bf 07} (2013) 147,
  [\href{http://arxiv.org/abs/1304.3803}{{\tt arXiv:1304.3803}}].

\bibitem{Langhoff:2020jqa}
K.~Langhoff and Y.~Nomura, {\it {Ensemble from Coarse Graining: Reconstructing
  the Interior of an Evaporating Black Hole}},  {\em Phys. Rev. D} {\bf 102}
  (2020), no.~8 086021, [\href{http://arxiv.org/abs/2008.04202}{{\tt
  arXiv:2008.04202}}].

\bibitem{Nomura:2020ewg}
Y.~Nomura, {\it {From the Black Hole Conundrum to the Structure of Quantum
  Gravity}},  \href{http://arxiv.org/abs/2011.08707}{{\tt arXiv:2011.08707}}.

\bibitem{Israel:1976ur}
W.~Israel, {\it {Thermo field dynamics of black holes}},  {\em Phys. Lett. A}
  {\bf 57} (1976) 107--110.

\bibitem{Takahasi:1974zn}
Y.~Takahasi and H.~Umezawa, {\it {Thermo field dynamics}},  {\em Collect.
  Phenom.} {\bf 2} (1975) 55--80.

\bibitem{Hartle:1976tp}
J.~Hartle and S.~Hawking, {\it {Path Integral Derivation of Black Hole
  Radiance}},  {\em Phys. Rev. D} {\bf 13} (1976) 2188--2203.

\bibitem{Hartle:1983ai}
J.~Hartle and S.~Hawking, {\it {Wave Function of the Universe}},  {\em Adv.
  Ser. Astrophys. Cosmol.} {\bf 3} (1987) 174--189.

\bibitem{lloyd2013pure}
S.~Lloyd, {\it {Pure state quantum statistical mechanics and black holes}},
  \href{http://arxiv.org/abs/1307.0378}{{\tt arXiv:1307.0378}}.

\bibitem{PhysRevLett.54.1350}
G.~Casati, B.~V. Chirikov, and I.~Guarneri, {\it {Energy-Level Statistics of
  Integrable Quantum Systems}},  {\em Phys. Rev. Lett.} {\bf 54} (Apr, 1985)
  1350--1353.

\bibitem{PhysRevLett.80.1373}
H.~Tasaki, {\it From quantum dynamics to the canonical distribution: General
  picture and a rigorous example},  {\em Phys. Rev. Lett.} {\bf 80} (Feb, 1998)
  1373--1376.

\bibitem{Stanford:2014jda}
D.~Stanford and L.~Susskind, {\it {Complexity and Shock Wave Geometries}},
  {\em Phys. Rev. D} {\bf 90} (2014), no.~12 126007,
  [\href{http://arxiv.org/abs/1406.2678}{{\tt arXiv:1406.2678}}].

\bibitem{Carmi:2017jqz}
D.~Carmi, S.~Chapman, H.~Marrochio, R.~C. Myers, and S.~Sugishita, {\it {On the
  Time Dependence of Holographic Complexity}},  {\em JHEP} {\bf 11} (2017) 188,
  [\href{http://arxiv.org/abs/1709.10184}{{\tt arXiv:1709.10184}}].

\bibitem{Susskind:2018pmk}
L.~Susskind, {\it {Three Lectures on Complexity and Black Holes}},  10, 2018.
\newblock \href{http://arxiv.org/abs/1810.11563}{{\tt arXiv:1810.11563}}.

\bibitem{CR}
J.~Chakravarty and S.~Raju, {\it {How do black holes manage to look larger from
  the inside than the outside?}},  {\em Unpublished Essay} (2020).

\bibitem{motl}
L.~Motl, {\it {One can't background-independently localize field operators in
  QG}},  {\em
  https://motls.blogspot.com/2013/08/one-cant-background-independently.html}
  (2013).

\bibitem{Itzhaki:2019cgg}
A.~Giveon and N.~Itzhaki, {\it {Stringy Information and Black Holes}},
  \href{http://arxiv.org/abs/1912.06538}{{\tt arXiv:1912.06538}}.

\bibitem{Ryu:2006bv}
S.~Ryu and T.~Takayanagi, {\it {Holographic derivation of entanglement entropy
  from AdS/CFT}},  {\em Phys. Rev. Lett.} {\bf 96} (2006) 181602,
  [\href{http://arxiv.org/abs/hep-th/0603001}{{\tt hep-th/0603001}}].

\bibitem{Ryu:2006ef}
S.~Ryu and T.~Takayanagi, {\it {Aspects of Holographic Entanglement Entropy}},
  {\em JHEP} {\bf 08} (2006) 045,
  [\href{http://arxiv.org/abs/hep-th/0605073}{{\tt hep-th/0605073}}].

\bibitem{Hubeny:2007xt}
V.~E. Hubeny, M.~Rangamani, and T.~Takayanagi, {\it {A Covariant holographic
  entanglement entropy proposal}},  {\em JHEP} {\bf 07} (2007) 062,
  [\href{http://arxiv.org/abs/0705.0016}{{\tt arXiv:0705.0016}}].

\bibitem{Lewkowycz:2013nqa}
A.~Lewkowycz and J.~Maldacena, {\it {Generalized gravitational entropy}},  {\em
  JHEP} {\bf 08} (2013) 090, [\href{http://arxiv.org/abs/1304.4926}{{\tt
  arXiv:1304.4926}}].

\bibitem{Faulkner:2013ana}
T.~Faulkner, A.~Lewkowycz, and J.~Maldacena, {\it {Quantum corrections to
  holographic entanglement entropy}},  {\em JHEP} {\bf 11} (2013) 074,
  [\href{http://arxiv.org/abs/1307.2892}{{\tt arXiv:1307.2892}}].

\bibitem{Engelhardt:2014gca}
N.~Engelhardt and A.~C. Wall, {\it {Quantum Extremal Surfaces: Holographic
  Entanglement Entropy beyond the Classical Regime}},  {\em JHEP} {\bf 01}
  (2015) 073, [\href{http://arxiv.org/abs/1408.3203}{{\tt arXiv:1408.3203}}].

\bibitem{Papadodimas:2017qit}
K.~Papadodimas, {\it {A class of non-equilibrium states and the black hole
  interior}},  \href{http://arxiv.org/abs/1708.06328}{{\tt arXiv:1708.06328}}.

\bibitem{Hamilton:2005ju}
A.~Hamilton, D.~N. Kabat, G.~Lifschytz, and D.~A. Lowe, {\it {Local bulk
  operators in AdS/CFT: A Boundary view of horizons and locality}},  {\em Phys.
  Rev. D} {\bf 73} (2006) 086003,
  [\href{http://arxiv.org/abs/hep-th/0506118}{{\tt hep-th/0506118}}].

\bibitem{Hamilton:2006fh}
A.~Hamilton, D.~N. Kabat, G.~Lifschytz, and D.~A. Lowe, {\it {Local bulk
  operators in AdS/CFT: A Holographic description of the black hole interior}},
   {\em Phys. Rev. D} {\bf 75} (2007) 106001,
  [\href{http://arxiv.org/abs/hep-th/0612053}{{\tt hep-th/0612053}}]. [Erratum:
  Phys.Rev.D 75, 129902 (2007)].

\bibitem{Roy:2015pga}
S.~R. Roy and D.~Sarkar, {\it {Hologram of a pure state black hole}},  {\em
  Phys. Rev. D} {\bf 92} (2015) 126003,
  [\href{http://arxiv.org/abs/1505.03895}{{\tt arXiv:1505.03895}}].

\bibitem{10.2307/79349}
M.~V. Berry and M.~Tabor, {\it Level clustering in the regular spectrum},  {\em
  Proceedings of the Royal Society of London. Series A, Mathematical and
  Physical Sciences} {\bf 356} (1977), no.~1686 375--394.

\bibitem{Milekhin:2020zpg}
A.~Milekhin, {\it {Quantum error correction and large $N$}},
  \href{http://arxiv.org/abs/2008.12869}{{\tt arXiv:2008.12869}}.

\bibitem{Aharony:2003sx}
O.~Aharony, J.~Marsano, S.~Minwalla, K.~Papadodimas, and M.~Van~Raamsdonk, {\it
  {The Hagedorn - deconfinement phase transition in weakly coupled large N
  gauge theories}},  {\em Adv. Theor. Math. Phys.} {\bf 8} (2004) 603--696,
  [\href{http://arxiv.org/abs/hep-th/0310285}{{\tt hep-th/0310285}}].

\bibitem{Sundborg:1999ue}
B.~Sundborg, {\it {The Hagedorn transition, deconfinement and N=4 SYM theory}},
   {\em Nucl. Phys. B} {\bf 573} (2000) 349--363,
  [\href{http://arxiv.org/abs/hep-th/9908001}{{\tt hep-th/9908001}}].

\bibitem{Witten:1998zw}
E.~Witten, {\it {Anti-de Sitter space, thermal phase transition, and
  confinement in gauge theories}},  {\em Adv. Theor. Math. Phys.} {\bf 2}
  (1998) 505--532, [\href{http://arxiv.org/abs/hep-th/9803131}{{\tt
  hep-th/9803131}}].

\bibitem{PhysRevLett.91.207901}
S.~Bose, {\it Quantum communication through an unmodulated spin chain},  {\em
  Phys. Rev. Lett.} {\bf 91} (Nov, 2003) 207901.

\bibitem{PhysRevA.71.032312}
M.~Christandl, N.~Datta, T.~C. Dorlas, A.~Ekert, A.~Kay, and A.~J. Landahl,
  {\it Perfect transfer of arbitrary states in quantum spin networks},  {\em
  Phys. Rev. A} {\bf 71} (Mar, 2005) 032312.

\bibitem{doi:10.1142/S0219749910006514}
A.~KAY, {\it Perfect, efficient, state transfer and its application as a
  constructive tool},  {\em International Journal of Quantum Information} {\bf
  08} (2010), no.~04 641--676,
  [\href{http://arxiv.org/abs/https://doi.org/10.1142/S0219749910006514}{{\tt
  https://doi.org/10.1142/S0219749910006514}}].

\bibitem{Ghosh:2017pel}
S.~Ghosh and S.~Raju, {\it {Loss of locality in gravitational correlators with
  a large number of insertions}},  {\em Phys. Rev. D} {\bf 96} (2017), no.~6
  066033, [\href{http://arxiv.org/abs/1706.07424}{{\tt arXiv:1706.07424}}].

\bibitem{Oxford}
G.~Akermann, J.~Baik, and P.~D. Francesco, {\em The Oxford Handbook of Random
  Matrix Theory}.
\newblock Oxford University Press, New York, 2011.

\bibitem{Gross:1980he}
D.~Gross and E.~Witten, {\it {Possible Third Order Phase Transition in the
  Large N Lattice Gauge Theory}},  {\em Phys.\ Rev.\ D} {\bf 21} (1980)
  446--453.

\bibitem{WADIA1980403}
S.~R. Wadia, {\it N = $\infty$ phase transition in a class of exactly soluble
  model lattice gauge theories},  {\em Physics Letters B} {\bf 93} (1980),
  no.~4 403 -- 410.

\bibitem{wadia2012study}
S.~R. Wadia, {\it A study of u(n) lattice gauge theory in 2-dimensions},
  \href{http://arxiv.org/abs/1212.2906}{{\tt arXiv:1212.2906}}.

\end{thebibliography}\endgroup

\end{document}